\documentclass[aps,prl,twocolumn,showpacs,floatfix,amsmath,superscriptaddress]{revtex4-1}

\usepackage{graphicx}
\usepackage{amsfonts}
\usepackage{color}

\newcommand{\mrm}[1]{\mathrm{#1}}                          

\newcommand{\vcrm}[1]{\boldsymbol{\mathrm{#1}}}    
\newcommand{\matx}[1]{\underline{#1}}                       
\newcommand{\unmx}[1]{\matx{\mathds{1}}}               

\newcommand{\imagi}{\mathrm{i}}                                
\newcommand{\ud}{\mathrm{d}}
\newcommand{\HH}{\mathcal{H}}

\newcommand{\sunit}[2]{#1~#2}                                    
\newcommand{\Angs}{\mbox{\r{A}}}
\renewcommand{\Re}{\mathrm{Re}}
\renewcommand{\Im}{\mathrm{Im}}

\makeatother

\begin{document}
\title{Renormalization of electron self-energies via their interaction with spin excitations: A first-principles investigation}
\author{Benedikt Schweflinghaus}
\affiliation{Peter Gr\"unberg Institut and Institute for Advanced Simulation, Forschungszentrum J\"ulich and JARA, 52425 J\"ulich, Germany}
\author{Manuel dos~Santos~Dias}
\affiliation{Peter Gr\"unberg Institut and Institute for Advanced Simulation, Forschungszentrum J\"ulich and JARA, 52425 J\"ulich, Germany}
\author{Antonio T.~Costa}
\affiliation{Department Instituto de Fisica, Universidade Federal Fluminense, Rio de Janeiro, Brazil}
\author{Samir Lounis}\email{s.lounis@fz-juelich.de}
\affiliation{Peter Gr\"unberg Institut and Institute for Advanced Simulation, Forschungszentrum J\"ulich and JARA, 52425 J\"ulich, Germany}
\date{\today}

\begin{abstract}
Access to magnetic excitation spectra of single atoms deposited on surfaces is 
nowadays possible by means of low-temperature inelastic scanning tunneling 
spectroscopy. We present a first-principles method for the calculation of 
inelastic tunneling spectra utilizing the Korringa-Kohn-Rostoker Green function 
method combined with time-dependent density functional theory and many-body 
perturbation theory. The key quantity is the electron self-energy describing 
the coupling of the electrons to the spin excitation within the adsorbate. 
By investigating  Cr, Mn, Fe and Co adatoms on a Cu(111) substrate, we 
spin-characterize the spectra and demonstrate that their shapes are altered by 
the magnetization of the adatoms, of the tip and the orbital decay into vacuum. 
Our method also predicts spectral features more complex than the steps obtained
by simpler models for the adsorbate (e.g., localized spin models).
\end{abstract}

\maketitle

\section{Introduction}

The study of magnetic properties of adatoms or clusters of few atoms deposited on surfaces is 
of crucial importance for the development of future magnetoelectronic devices that push the boundaries of efficiency 
with respect to both density of binary information and temporal stability. In nanospintronics, spin and charge currents can be strongly affected by the scattering 
of electrons by collective excitations, such as spin excitations (SE)~\cite{Fabian+RMP04}. 
The effect of such scattering can be described with an electronic self-energy. 
Besides its impact in nanotechnologies, the interaction between electrons 
and SE ($I_{\mathrm{e-SE}}$) is a fundamental issue. It can have strong impact on 
spin-fluctuations~\cite{MoriyaSpringerSeries85}, superconductivity in 
Fe-pnictides~\cite{Mazin+PRL08, DagattoRMP94}, and dynamics of atomic-scale 
magnets~\cite{Khajetoorians+Science13}.

In angle-resolved photoemission spectroscopy, the $I_{\mathrm{e-SE}}$ shows up as a kink in the 
band-structure~\cite{Schaefer+PRL04,Cui+JMM07,Hofmann+PRL09}, while for low-temperature inelastic 
scanning tunneling spectroscopy (ISTS) the signature of the 
SE is found in the conductance~\cite{Heinrich+Science04,Hirjibehedin+Science06,
Balashov+PRL09,Khajetoorians+PRL11,Chilian+PRB11,Otte+PRL13,Khajetoorians+PRL13}. 
In ISTS of nanostructures deposited on surfaces, the electrons interact with the substrate during the 
tunneling process and exchange energy and possibly spin angular momentum. This leads to additional 
tunneling channels usually assumed to manifest as a steplike increase of the conductance. 

The nature of both the adsorbate and substrate is of primordial importance in ISTS. 
Indeed, hybridization between their respective 
electronic states plays a major role in defining the main characteristics of the SE 
spectra~\cite{MillsLedererPR67,MunizMillsPRB03,Lounis+PRL10,Lounis+PRB11}, such as 
excitation energies and lifetimes. Recently, it was shown~\cite{Lounis+PRL10,Lounis+PRB11,Khajetoorians+PRL11,Chilian+PRB11,Khajetoorians+PRL13} that the imaginary part of the 
transverse dynamical magnetic susceptibility, $\chi$, calculated from first principles, 
can be used to reliably extract the density of SE states that contains and explains the 
previously mentioned characteristics, albeit it does not provide theoretical inelastic spectra. 
Also, we note that several model calculations based on a Heisenberg Hamiltonian~\cite{Hirjibehedin+Science06,LorenteGauyacqPRL09,PerssonPRL09,Fernandez-RossierPRL09,FranssonNL09} or 
beyond~\cite{Hurley+PRB11,Hurley+PRB12} were proposed to understand the ISTS spectra. 
However, they often rely on a fitting procedure of experimental input.

Although a tremendous effort has been made in the investigation of SE, many questions remain open, for instance 
the asymmetry of the inelastic spectra, the nonobservation of SE while a Heisenberg model 
will always predict their presence, the spin-nature of the observed inelastic spectra. 
The goal of this article is to answer some of them. 
We present a first-principles method, based on the Korringa-Kohn-Rostoker Green function (KKR-GF) method embedded in a time-dependent density functional theory (TDDFT) formalism in combination with many-body perturbation theory (MBPT), 
which allows a realistic description of theoretical inelastic tunneling spectra. 
The advantage of such a scheme lies in the direct access to the Green function renormalized by the presence of $I_{\mathrm{e-SE}}$. 
Thus, we extract the related self-energies and their impact on the electronic structure. 
This enables the calculation of realistic excitation spectra in the vacuum above the impurity that are comparable to ISTS measurements, in the spirit of the Tersoff-Hamann approximation~\cite{TersoffHamannPRL83}. 
We explain many of the experimental observations that are not understood by demonstrating that: 
(i) the usual asymmetry in the inelastic spectra is induced by the magnetization of the adsorbate and of the ISTS tip, 
(ii) the shape of the SE signature is not necessarily a step in the conductance, 
(iii) additional spectroscopic features induced by the $I_{\mathrm{e-SE}}$ are found, and 
(iv) the spin-character of the excitation signature is 
revealed. 
After a brief discussion of our scheme we analyze results for single 3$d$ adatoms deposited on a Cu(111) surface and compare our simulations for Fe to available measurements~\cite{Khajetoorians+PRL11}.

\section{Method}

\begin{figure}[b]
 \centering
  \includegraphics[scale=1]{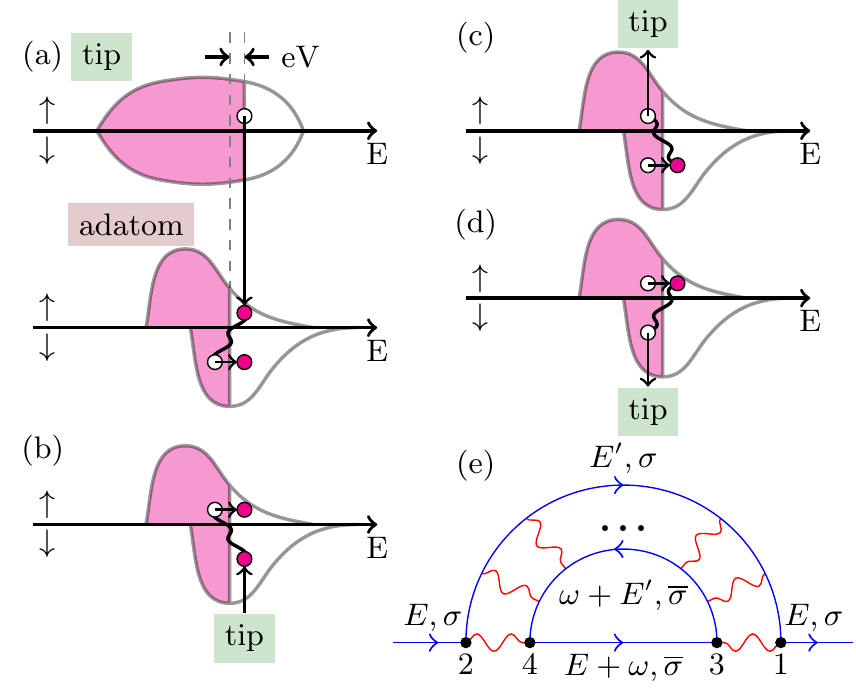}\vspace{-1em}
 \caption{(Color online) Four basic spin-flip processes superimposed on schematic spin-resolved DOS for the adatom ((a)-(d)) and the tip (only (a)). 
Electrons are indicated by filled red circles, holes by empty white circles. 
The wiggly lines represent e-h pair interactions. 
The probability associated with each process depends on both the sign and the magnitude of the applied bias voltage $V$. 
All four processes contribute to the Feynman diagram (e), incorporating the infinite series of interactions (see Eqs.~(\ref{eq:SelfEnergy}) and (\ref{eq:DysonLikeEq}) and related discussion).}
 \label{fig:Feynman_Selfe}
\end{figure}

The self-energy of interest, $\Sigma$, describes spin-flip processes as visualized, for simplicity, in Fig.~\ref{fig:Feynman_Selfe}: 
for example, an electron with spin up travels from the tip to the surface where it excites an electron in the 
minority band [Fig.~\ref{fig:Feynman_Selfe}(a)]. 
The hole created in the minority-spin channel and the {\it tunneling} electron can form an \textit{e-h} pair of 
opposite spins. 
Other processes are obtained by swapping the spin labels ([Fig.~\ref{fig:Feynman_Selfe}(b)], the role of particles and holes [Fig.~\ref{fig:Feynman_Selfe}(c)], or both [Fig.~\ref{fig:Feynman_Selfe}(d)]. 
All four processes can be subsumed under the Feynman diagram as given in Fig.~\ref{fig:Feynman_Selfe}(e). 
The \textit{e-h} pairs after renormalization via the mediating interactions (wiggly lines) lead to correlated 
spin-flip excitations (magnons in extended systems). 
Processes in Figs.~\ref{fig:Feynman_Selfe}(a) and \ref{fig:Feynman_Selfe}(c) contribute to $\Sigma^{\uparrow}$ while 
$\Sigma^{\downarrow}$ is determined by processes in Figs.~\ref{fig:Feynman_Selfe}(b) and \ref{fig:Feynman_Selfe}(d). 
Depending on the electronic structure, as exemplified in Fig.~\ref{fig:Feynman_Selfe}, some processes can be dominant. 
This is related to the density of states (DOS) for electrons and holes available for the different processes. 
In Fig.~\ref{fig:Feynman_Selfe}(b), the amplitude of the \textit{e-h} pair defined by the 
unoccupied minority-spin states and occupied majority-spin states is much larger than the 
amplitude of the \textit{e-h} pair defined by the occupied minority-spin states and unoccupied majority-spin 
states shown in Fig.~\ref{fig:Feynman_Selfe}(d). 
Thus, one expects the self-energy for the majority-spin channel to be mainly shaped by the 
process in Fig.~\ref{fig:Feynman_Selfe}(c) while the minority-spin channel would be mainly shaped by the 
process in Fig.~\ref{fig:Feynman_Selfe}(b), as intuitively proposed in Ref.~[\onlinecite{Balashov+PRB08}]. 

The object consisting of the two half-circles connecting the points 2 with 1 
and 3 with 4 interacting via the wiggling line resembles that of the transverse dynamical magnetic susceptibility. 
We note that the diagram in Fig.~\ref{fig:Feynman_Selfe}(e) is one of the many that, 
in the $T$-approximation~\cite{BaymKadanoffPR61,KanamoriPTP63}, describe the renormalization of 
the mediating interaction $U$ to the scattering $T$-matrix via $\Sigma$~\footnote{$\Sigma$ is given as a convolution $G(T-U)$. 
 Since $T = U + U\chi U$, $\Sigma$ simplifies to $GU\chi U$ where $\chi$ is the transverse magnetic response function.}. 
Naturally, such Feynman diagrams induced by $I_{\mathrm{e-SE}}$ were investigated for decades with simple models (see, {\it e.g.},~Refs.~[\onlinecite{WangScalapinoPR68}] and [\onlinecite{AppelbaumBrinkmanPR69}]). 
Hertz and Edwards~\cite{EdwardsHertzJPF73-I,EdwardsHertzJPF73-II}, for instance, devised a scheme to avoid self-consistent calculations with the computed self-energy (see also Ref.~[\onlinecite{CelascoCorriasNC76}]).
Recently, realistic models based on the evaluation of $T$ using either a tight-binding scheme~\cite{HongMillsPRB99} or density functional theory in the $GW$ approximation~\cite{Zhukov+PRL04, Zhukov+PRB06, MuellerMasterThesis11} were developed and applied for bulk materials~\footnote{In Refs.~[\onlinecite{Zhukov+PRL04,Zhukov+PRB06,MuellerMasterThesis11}], the screened interaction, $W$, as calculated in $GW$ was used in evaluating $T$ instead of $U$. Romaniello \textit{et al}.~\cite{Romaniello+PRB12} discusses the different forms of $T$ depending on the strength of screening.}. 

The Feynman Diagram in Fig.~\ref{fig:Feynman_Selfe}(e) is translated to the following form considering a local and adiabatic approximation for $U$ ($\sigma =\, \uparrow,\downarrow$ and $\overline\sigma =\, \downarrow,\uparrow$):
\begin{eqnarray}
\Sigma^\sigma(\vcrm{r}_1,\vcrm{r}_2;E)
&=& -\frac{U(\vcrm{r}_1)\, U(\vcrm{r}_2)}{\pi}\times \nonumber \\
& & \hspace{-7.5em} \times\bigg[\int_0^\infty \mathrm{d}\omega\, \Im\left[ G_0^{\overline\sigma}(\vcrm{r}_1,\vcrm{r}_2;\omega+E) \chi^{\sigma\overline\sigma}(\vcrm{r}_1,\vcrm{r}_2;\omega) \right] \nonumber \\
& & \hspace{-7.5em} - \int_0^{E_\mathrm{F}-E} \hspace{-1em} \mathrm{d}\omega\, \Im\left[ G_0^{\overline\sigma}(\vcrm{r}_1,\vcrm{r}_2;\omega+E) \right] \chi^{\sigma\overline\sigma}(\vcrm{r}_2,\vcrm{r}_1;\omega)^* \bigg] \; .
\label{eq:SelfEnergy}
\end{eqnarray}
Here, $\chi^{\uparrow\downarrow}$ and $\chi^{\downarrow\uparrow}$ correspond, respectively, to $\chi^{+-}$ and $\chi^{-+}$; see Appendix~A. 
Since we are interested in simulating ISTS-related experiments we can proceed to the change of variables: 
$E=E_\mathrm{F}+V$ and $V$ corresponding to the applied bias voltage. 

$\chi^{\sigma\overline\sigma}$ is the transverse dynamical magnetic susceptibility that can be calculated from the Dyson-like equation as given in a matrix notation:
\begin{eqnarray}
\chi^{\sigma\overline\sigma} = \chi^{\sigma\overline\sigma}_0 + \chi^{\sigma\overline\sigma}_0 U \chi^{\sigma\overline\sigma} \; .
\label{eq:DysonLikeEq}
\end{eqnarray}
Within TDDFT, which is the basis of this work, $\chi^{\sigma\overline\sigma}_0$ is the response function of the Kohn-Sham system, which is connected to the full susceptibility via the exchange and correlation kernel, $U$, which simplifies in the adiabatic local density approximation (ALDA) to $U(\vcrm{r}) = \frac{B^\mathrm{xc}(\vcrm{r})}{m(\vcrm{r})}$ (see, {\it e.g.},~Refs.~[\onlinecite{Lounis+PRL10}] and [\onlinecite{Lounis+PRB11}]). 
Equation~(\ref{eq:DysonLikeEq}) also occurs in many-body perturbation theory (MBPT), in the random-phase approximation (RPA)~\cite{MillsLedererPR67,MunizMillsPRB03,Sasioglu+PRB10}. 
There, $\chi_0$ is the noninteracting susceptibility that connects to the full susceptibility via $U$, the screened Coulomb interaction. 
It was already shown that a mapping between the two schemes is possible by considering $U$ as the exchange and correlation kernel~\cite{Lounis+PRL10,Lounis+PRB11}. 
A similar connection in the spirit of the Bethe-Salpeter equation was proposed for the case of charge excitations~\cite{KarlssonAryasetiawanIJMPB04} or for SE~\cite{Brandt71}. 
Our strategy is thus to use TDDFT to extract the susceptibility. 
Once $\Sigma$ is known, we plug it into the Dyson equation given in a matrix notation $G = G_0 + G_0 \Sigma G$ with the Green function $G_0$ containing the reference electronic structure.

\begin{figure*}[bth]
 \centering
  \includegraphics[trim = 0mm 88mm 0mm 0mm, clip, width=0.9\textwidth]{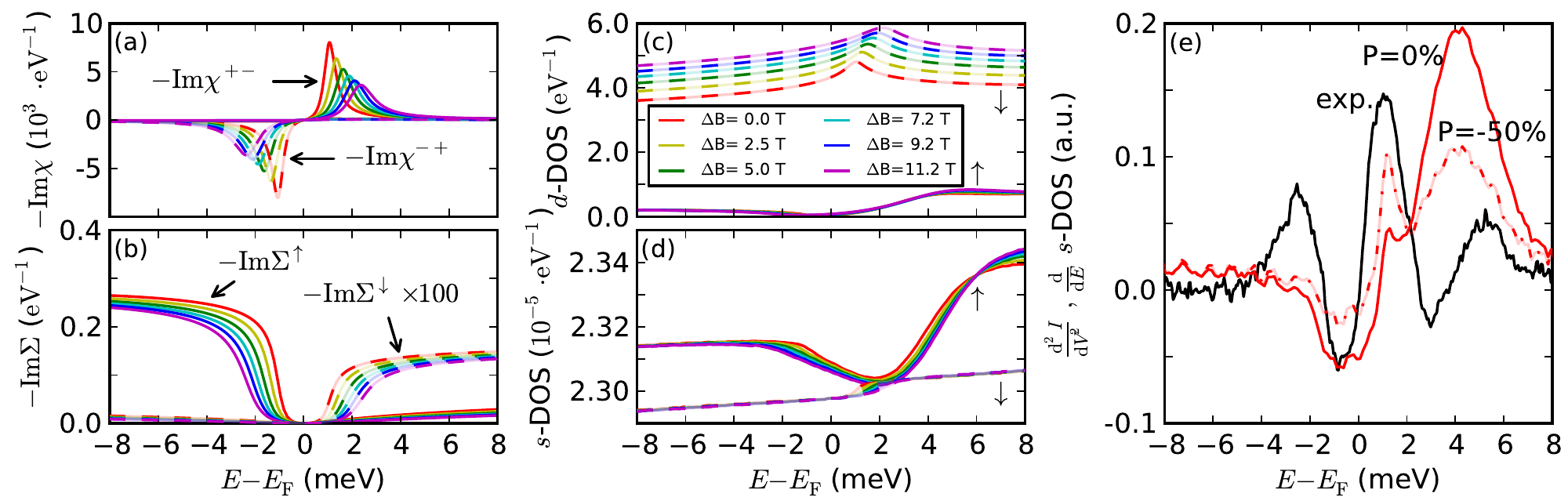}\vspace{-1em}
 \caption{(Color online) The energy dependence of the spin-resolved key quantities 
for Fe adatoms on Cu(111): (a) the response function, (b) the self-energy, 
(c) the $d$-DOS of the adatom, and (d) $s$-DOS in vacuum, for different magnetic fields. 
All quantities are plotted for spin-up (solid lines, $\uparrow$) and spin-down (dashed lines, $\downarrow$). 
(e) A comparison to the derivative of the conductance spectra obtained in experiment (see Ref.~[\onlinecite{Khajetoorians+PRL11}]). The agreement improves when instead of a nonpolarized tip (solid red curve) a polarization of $P=\sunit{-50}{\%}$ is assumed (dashed red curve); see Appendix~E for a detailed discussion.}
 \label{fig:results_Susc_Selfe_DOS}
\end{figure*}

\section{Results and Discussion}

In order to mimic the effect of spin-orbit coupling, we apply an auxiliary external magnetic field with $\mu_\mrm{B}B_0 \sim$ \sunit{0.5}{meV} that opens a gap in the excitation spectra at the Larmor resonance frequency, $\omega_\mathrm{res}=\mu_\mrm{B} g B_0$ ($g \sim 2$ is the Land\'e factor), which matches the experimental data for the Fe adatom~\cite{Khajetoorians+PRL11}. 
For the sake of comparison, the same auxiliary field is used for all adatoms. 

For the electronic structure, use is made of the KKR-GF method~\cite{KKR} in the 
atomic sphere approximation (ASA) with full charge density in the local spin density 
approximation, as parametrized by Vosko, Wilk, and Nusair~\cite{VoskoWilkNusairCJP80}. 
A slab of 22 Cu layers stacked in the (111) direction augmented by two vacuum regions was used to define the 
undisturbed Cu(111) surface, using the experimental lattice constant ($a=\sunit{3.615}{\Angs}$). 
From this surface a real space cluster is cut out surrounding the position to be occupied by the impurity adatom. 
A relaxation of the adatom by 14 \% towards the surface was considered (0 \% corresponds to the ideal interlayer separation in bulk, $a/\sqrt{3}=\sunit{2.087}{\Angs}$).

We analyze the spin excitations of several transition metal adatoms on a Cu(111) slab with 22 Cu layers. 
To calculate $\chi^{\sigma\overline\sigma}$ we consider the response of the systems to a site- and frequency-dependent transverse magnetic field where a projection to a localized basis set is considered ($d$-wave functions defined at $E_\mrm{F}$). 
For more details see Refs.~[\onlinecite{Lounis+PRL10}] and [\onlinecite{Lounis+PRB11}]. 
In this scheme, the transverse susceptibility simplifies to a single number for a single adatom (spherical approximation), which is reasonable since for most of the adatoms magnetic moments is carried by $d$ electrons. 
$\Im\chi^{\sigma\overline\sigma}$ for different magnetic fields $\Delta B=B-B_0$ are shown in Fig.~\ref{fig:results_Susc_Selfe_DOS}(a) for the Fe adatom.

We proceed by discussing the imaginary part of the self-energy projected on the $d$ basis and integrated within the atomic sphere surrounding the adatom,
\begin{eqnarray}
\Im\Sigma_{mm'}^{\sigma}(E_\mathrm{F}+V)
&=& \nonumber \\
& & \hspace{-32mm} -U^2 \int_{0}^{-V}\mathrm{d}\omega\ n_{mm'}^{\overline\sigma}(E_\mathrm{F}+V+\omega)\cdot\Im\left[\underline\chi^{\sigma \overline\sigma}(\omega)^*\right] \; ,
\label{eq:proj_ImSelfe}
\end{eqnarray}
where $n^{\sigma}(E)=-1/\pi \cdot \Im G_0^{\sigma}(E)$ is the local DOS obtained for the initial Green function, $\underline\chi$ is the spherical part of the susceptibility, {\it i.e.}, $\underline\chi=\sum_{mm'}\chi_{mm';m'm}$ and $m$, $m'$ label the $d$-orbitals. 
If one considers $n(E)$ to be featureless, an energy integration of $\Im\chi^{\sigma\overline\sigma}$ is performed in Eq.~(\ref{eq:proj_ImSelfe}). 
Naturally, one expects a steplike function as soon as the integration goes over a bias voltage $V$ equal to  $\omega_{\mathrm{res}}$. 
The resulting spin-resolved self-energy is shown in Fig.~\ref{fig:results_Susc_Selfe_DOS}(b), where the trace of $\Sigma^\uparrow$ and  $\Sigma^\downarrow$ are indicated by solid and dashed lines, respectively. 
Because of the relation between the step positions and $\omega_{\mathrm{res}}$, the gap between them increases with 
$B$. 
Whereas the height of the resonances in $\chi^{\sigma\overline\sigma}$ are equal with 
respect to the two spin channels, the step height in the self-energy differs by a factor of 
about 100. 
This can be understood as the resonance being weighted by the DOS of the opposite spin channel, \textit{cf}.~Eq.~(\ref{eq:proj_ImSelfe}): 
if there is only a small number of $\overline\sigma$-states available, the scattering is unlikely to happen. 
In contrast to the extremely small $n_\mrm{Fe}^\uparrow$, the $n_\mrm{Fe}^\downarrow$ displays a large resonance; 
see Appendix~B. 
Since the step widths are related to the line widths extracted from the susceptibility 
peaks, they increase when the excitation energy $\omega_\mathrm{res}$ increases. 

The results for the $d$ orbitals of the Fe adatom and for the $s$ orbitals of the vacuum site are shown in Figs.~\ref{fig:results_Susc_Selfe_DOS}(c) and \ref{fig:results_Susc_Selfe_DOS}(d), respectively. 
Whereas the self-energy shows a height difference between the two spin channels of about two orders of magnitude, the resulting DOS magnitude do not differ much anymore. 
Although in the adatom, the $d$-DOS for the minority-spin channel is larger than the one for the majority-spin channel 
(because of the large minority-spin resonance), the opposite is found in vacuum for the $s$-DOS. 
An analysis of the orbital contributions to the total adatom-DOS, for instance the orbitals extending farthest to vacuum, shows 
that the spin asymmetry within the adatom is orbital-dependent. Indeed, contrary to the $d_{z^2}$ and $s$ state, the $p_z$ states have majority-spin DOS larger than the 
minority-spin DOS, similar to the spin asymmetry in the vacuum; see Appendix~C. Hybridization, interferences effects, and decays of orbitals shapes the final form of 
the vacuum DOS. For example, the peaklike feature in the minority-spin channel of the $d$ orbital at the Fe adatom (see solid lines) can evolve into a steplike feature for the $s$ states at the vacuum site, which in the presented calculations is about \sunit{6.3}{\r{A}} above the adatom.

In Fig.~\ref{fig:results_Susc_Selfe_DOS}(e), we show a comparison of the experimental $\mrm{d}^2I/\mrm{d}V^2$ data for an Fe adatom from Ref.~[\onlinecite{Khajetoorians+PRL11}] with the energy derivative of our $s$-DOS in vacuum, \textit{cf.}~Fig.~\ref{fig:results_Susc_Selfe_DOS}(d). 
The experimental spectrum shows two distinct sets of features. Since the shape of the SE signature is not a perfect step in the conductance, 
the first derivative leads to  peak and dip pairs at $\sunit{\pm1}{\mathrm{meV}}$ and $\sunit{\pm3}{\mathrm{meV}}$; 
see Appendix~D. 
The shape of the SE signature in $s$-DOS is slightly different from the experimental ones, explaining the 
absence of the dip at $\sunit{-3}{\mathrm{meV}}$ in the corresponding energy derivative. 
Interestingly, there is an additional peak in the experimental spectrum at $\sunit{+5}{\mathrm{meV}}$ that has 
no matching dip at $\sunit{-5}{\mathrm{meV}}$ in good agreement with our simulations (satellite at $\sunit{+4}{\mathrm{meV}}$). 
The origin of this extra feature can be traced back to $\Re\Sigma$. 
In the expression for $G$, the denominator $(1-G_0\Sigma)$ causes a resonance when $\Im(G_0\Sigma) \ll 1$ and 
$\Re(G_0\Sigma)$ is close to 1. 
This condition seems to be satisfied in the majority-spin channel around $E_\mathrm{F}$.  
The self-energy thus is acting as an additional potential on the electrons, which can lead to 
satellites very similar to split-off states observed when adatoms interact with surface states~\cite{Lounis+PRB06,Limot+PRL05}. 
We studied the effect of the spin-polarization of the tip with a simple model, see Appendix~E, choosing $P=\sunit{0}{\%}$ or $P=\sunit{-50}{\%}$; the overall shape of the spectrum can be modified by changing the weight of the spin-resolved SE signature [\textit{cf.}~solid versus dashed red line in Fig.~\ref{fig:results_Susc_Selfe_DOS}(e)]. 
This can improve the agreement with the experiment and indicates that the shape of the inelastic spectra is not only a function of the adsorbate but also of the polarization of the tip. 

\begin{figure}[tb]
 \centering
   \includegraphics[width=0.45\textwidth]{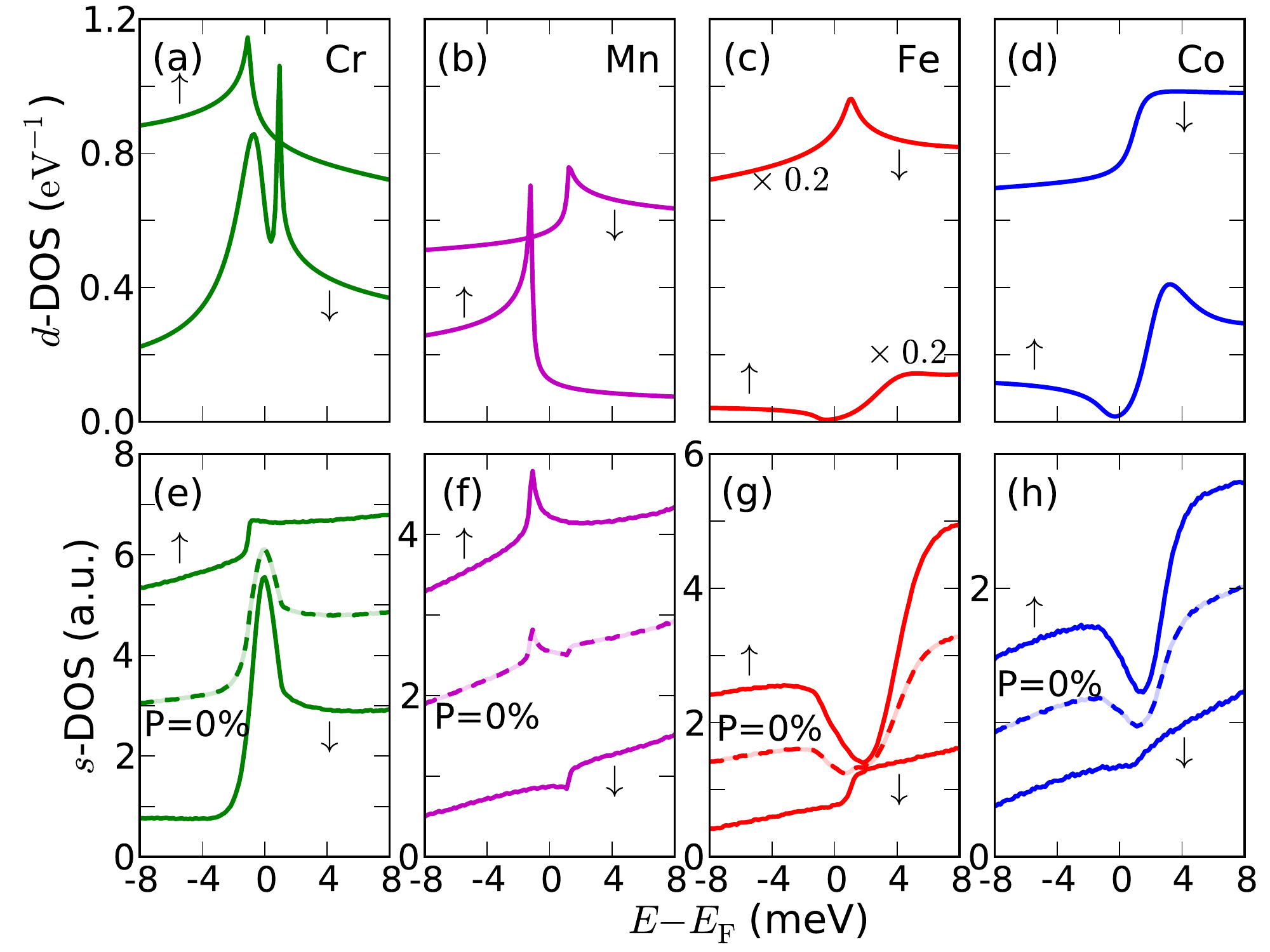}\vspace{-1em}
 \caption{(Color online) DOS renormalized by spin excitations for the four 3$d$ adatoms. 
 Solid lines represent spin-resolved DOS and dashed lines represent the respective 
 spin-average ($\frac{\uparrow+\downarrow}{2}$), relevant for a nonpolarized tip, $P=\sunit{0}{\%}$; see Appendix~E. 
 Top row: DOS for the $d$ orbitals of the impurity atom. 
 Bottom row: DOS for vacuum $\sim$\sunit{6.3}{\r{A}} above the adatoms ($s$ orbitals).}
 \label{fig:DOS_TM_vac}
\end{figure}

The excitation spectra of Co, Mn, and Cr adatoms are given in Fig.~\ref{fig:DOS_TM_vac}. 
The top row [Figs.~\ref{fig:DOS_TM_vac}(a)-\ref{fig:DOS_TM_vac}(d)] shows the DOS for the spin-resolved adatoms $d$ orbitals and the bottom row [Figs.~\ref{fig:DOS_TM_vac}(e)-\ref{fig:DOS_TM_vac}(h)] shows 
the spin-resolved and the spin-averaged (dashed lines) vacuum $s$ orbitals above the impurity. 
The Co adatom's spectrum reveals some similarities to those of the Fe adatom. 
For the majority-spin their shapes, including the additional satellite, are nearly identical. 
For the minority-spin channel, however, the SE feature almost vanishes in vacuum. In contrast to Fe and Co, the Mn renormalized DOS do not show additional satellites. 
However, the excitation signatures are steplike functions with a peaklike resonance at the edges. For the Cr adatom, peaklike structures are observed in the $d$-DOS, which transform in vacuum into a reversed step for the majority-spin channel, while in the minority-spin channel the SE and the satellite overlap at $E_\mathrm{F}$.

We note that Co adatoms on Cu(111) is a traditional Kondo system and that processes leading to Kondo behavior are not included in our scheme~\footnote{But spin excitations and Kondo can coexist~\cite{Otte+NP08}.}. In contrast to Co, Fe shows no Kondo signature down to \sunit{0.3}{K}~\cite{Khajetoorians+PRL11}. 
This is strengthened by the measurements of magnetic exchange interactions among Fe adatoms~\cite{Khajetoorians+NP12}. Cr and Mn adatoms on Cu(111) are expected to behave as on Au(111)~\cite{Jamneala+PRB00}, where no Kondo behavior is observed.

The lifetime of the SE, $\tau_\chi$, is given by the line width of $\Im\chi$, which is different from the lifetime extracted from the inelastic spectra, $\tau_{\mathrm{DOS}}$. 
Both lifetimes are calculated from $\tau=\hbar / 2 \Delta E$, where $\Delta E$ is the full-width half-maximum of the signature of the spin excitations. 
Because of the convolution with the one-electron GF's, more information is encoded in  $\tau_{\mathrm{DOS}}$, which is the only quantity reachable experimentally. 
Contrary to $\tau_{\chi}$, $\tau_{\mathrm{DOS}}$ is spin-dependent and the difference between the spin channels can reach a factor 5. 
Indeed, $\tau_{\chi}=\{1.9,2.9,0.6,0.2\}$~ps for, respectively, \{Cr, Mn, Fe, and Co\}, while the sequence changes to $\{1.1,0.5,0.1,0.3\}$~ps for $\tau_{\mathrm{DOS}}^{\uparrow}$ and $\{0.4,0.8,0.5,0.1\}$~ps for $\tau_{\mathrm{DOS}}^\downarrow$. 
Furthermore, for some systems the additional satellite contribute to the effective lifetime of the excitation signature (Cr is the extreme case). 
The lifetimes of Co and Fe adatoms are up to one order of magnitude smaller than those of Mn and Cr 
adatoms when the resonance of the susceptibility is used. 
This is due to the relatively small minority-DOS at $E_\mathrm{F}$ for the latter two systems: the excited electron cannot easily find an unoccupied state to deexcite to and thus the excitation lifetime is longer; see Fig.~\ref{fig:Feynman_Selfe}.

\section{Conclusion}

In summary, a first-principles approach to inelastic magnetic excitation spectra is developed utilizing the KKR-GF method combined with TDDFT and MBPT. 
We illustrate its capabilities by investigating 3$d$ adatoms on a Cu(111) surface with a focus on Fe impurities. 
We relate the asymmetry of the inelastic spectra (height and lifetime) to the electronic and magnetic structure of the adatom as well as the magnetization of the ISTS tip. 
The spin-character of the excitations above and below $E_\mrm{F}$ is explained. 
Most importantly, the spectra can have different shapes, including a steplike form, and extinction of the signature of the excitations can occur. 
Also, nontrivial spectral satellites are obtained, which we believe to be observable experimentally and could even be mistaken as being the signature of SE. 
Further work involves handling spin-orbit coupling, the effects of self-consistency on the self-energy, and approximations beyond the ALDA. 

\section*{Acknowlegdments}

We acknowledge the contributions of late D.~L.~Mills. 
Also we thank S.~Bl\"ugel, E.~\c{S}a\c{s}\i o\u{g}lu, P.~H.~Dederichs, A.~A.~Khajetoorians, and J.~Wiebe for fruitful discussions. 
This work is supported by the HGF-YIG Programme VH-NG-717 (Functional Nanoscale Structure and Probe Simulation Laboratory, Funsilab).

\begin{appendix}

\section{The Kohn-Sham susceptibility and the self-energy}

In the definition of the Kohn-Sham susceptibility used in Eq.~(1) in the main text we follow Lounis \textit{et al.}~\cite{Lounis+PRL10,Lounis+PRB11}.

The Kohn-Sham (KS) Green function (GF) is the resolvent of the corresponding Hamiltonian, $G_{\text{KS}}(E) = (E - \mathcal{H}_{\text{KS}})^{-1}$. 
In the Korringa-Kohn-Rostoker Green function (KKR-GF) method, space is partitioned into nonoverlapping regions surrounding the atoms, labeled $i$. 
These regions are taken as spherical in the atomic sphere approximation (ASA), and the KS potential is also assumed to be spherical around each atom, $V^{\text{KS}}_i(r)$, with $r = |\vec{r}\,|$ and $\hat{r} = \vec{r}/r$. 
Then the KS GF is expressed in terms of energy-dependent scattering solutions for each atomic potential, $R_{i\ell}^\sigma(r;E)\,Y_{L}(\hat{r})$ and $H_{i\ell}^\sigma(r;E)\,Y_{L}(\hat{r})$, which are products of radial functions and (real) spherical harmonics, for each spin $\sigma = \;\uparrow,\,\downarrow$ and angular momentum $L = (\ell,m)$. 
$R_{i\ell}^\sigma(r;E)$ is regular at the center of the ASA sphere, and $H_{i\ell}^\sigma(r;E)$ diverges there. 
The KKR-GF then takes the form
\begin{eqnarray}
  G_{ij}^\sigma(\vec{r}\,,\vec{r}\,';E) &=& \sum_{LL'}Y_L(\hat{r}) \big(\delta_{ij}\sqrt{E}\,R_{i\ell}^\sigma(r_<;E)\,H_{i\ell}^\sigma(r_>;E) \nonumber\\
  &&\hspace{-4em} + R_{i\ell}^\sigma(r;E)\,G^\sigma_{iL,jL'}(E)\,R_{j\ell'}^\sigma(r';E)\big) Y_{L'}(\hat{r}') \; ,
\end{eqnarray}
where $r_< = \min(r,r')$ and $r_> = \max(r,r')$, and $G^\sigma_{iL,jL'}(E)$ is the structural GF, describing backscattering effects.\\

As explained in Refs.~[\onlinecite{Lounis+PRL10}] and [\onlinecite{Lounis+PRB11}], near the Fermi energy ($E_\text{F}$) one may approximate $R_{i\ell}^\sigma(r;E) \approx R_{i\ell}^\sigma(r;E_\text{F})$.
Furthermore, given that the states of interest are the $d$ orbitals of a single magnetic adatom, one may drop the site label $i$ and keep only $\ell = 2$, projecting on the regular scattering solutions computed at $E_\text{F}$:
\begin{eqnarray}
  G_{d,mm'}^\sigma(E) &=& \!\int\!\!\text{d}\vec{r}\,\!\int\!\!\text{d}\vec{r}\,'\,R_{d}^\sigma(r;E_\text{F})\,Y_{2m}(\hat{r}) \times \nonumber\\
                      & & \hspace{-2em} \times G^\sigma(\vec{r}\,,\vec{r}\,';E)\,R_{d}^\sigma(r';E_\text{F})\,Y_{2m'}(\hat{r}')
\end{eqnarray}
This defines the projection on the $d$ orbitals of the adatom of the KKR-GF, upon suitable normalization.\\

The transverse magnetic KS susceptibility is given in terms of the KS GFs as


\begin{eqnarray}
  \chi^{\sigma\bar{\sigma}}_{0,ij}(\vec{r}\,,\vec{r}\,';\omega)
    &=& -\frac{1}{\pi}\!\int^{E_\text{F}}\!\!\!\!\text{d}E \nonumber \\
    & & \hspace{-6em}\Big(G_{ij}^{\bar\sigma}(\vec{r}\,,\vec{r}\,';E+\omega+\imagi 0)\,\text{Im}\,G_{ji}^{\sigma}(\vec{r}\,',\vec{r}\,;E) + \nonumber \\
    & & \hspace{-6em}+ \text{Im}\,G_{ij}^{\bar\sigma}(\vec{r}\,,\vec{r}\,';E)\,G_{ji}^{\sigma}(\vec{r}\,',\vec{r}\,;E-\omega-\imagi 0)\Big) \, .
\end{eqnarray}


Here, $\chi^{\uparrow\downarrow}$ and $\chi^{\downarrow\uparrow}$ correspond to $\chi^{+-}$ and $\chi^{-+}$, respectively. 
Introducing the projection on the $d$ orbitals, this leads to

\begin{widetext}

\begin{eqnarray}
  \chi^{\sigma\bar{\sigma}}_{0d}(\vec{r}\,,\vec{r}\,';\omega) &=& \nonumber\\
  & &\hspace{-5em} \!\!\!\!\sum_{m_1m_2m_3m_4}\!\!\!\! R_{d}^{\bar\sigma}(r;E_\text{F})\,Y_{2m_1}(\hat{r})\,R_{d}^{\bar\sigma}(r';E_\text{F})\,Y_{2m_2}(\hat{r}')\,R_{d}^{\sigma}(r';E_\text{F})\,Y_{2m_3}(\hat{r}')R_{d}^{\sigma}(r;E_\text{F})\,Y_{2m_4}(\hat{r})\,\chi^{\sigma\bar{\sigma}}_{0d,m_1m_2m_3m_4}(\omega) \; ,
\end{eqnarray}

\end{widetext}

where
\begin{eqnarray}
  \chi^{\sigma\bar{\sigma}}_{0d,m_1m_2m_3m_4}(\omega)
  &=& -\frac{1}{\pi}\!\int^{E_\text{F}}\!\!\!\!\text{d}E\, \times \nonumber\\
  & & \hspace{-6em} \Big(G_{d,m_1m_2}^{\bar\sigma}(E+\omega+\imagi 0)\,\text{Im}\,G_{d,m_3m_4}^{\sigma}(E) + \nonumber\\
  & & \hspace{-6em} + \text{Im}\,G_{d,m_1m_2}^{\bar\sigma}(E)\,G_{d,m_3m_4}^{\sigma}(E-\omega-\imagi 0)\Big) \; .
\end{eqnarray}\\
At this stage it is useful to recall the magnetization sum rule, see Refs.~[\onlinecite{Lounis+PRL10}] and [\onlinecite{Lounis+PRB11}],
\begin{eqnarray}
  m_i(\vec{r}\,)
    &=& \sum_j\!\int\!\!\text{d}\vec{r}\,'\,\chi^{\uparrow\downarrow}_{0,ij}(\vec{r}\,,\vec{r}\,';0)\,B_{\text{xc},j}(\vec{r}\,') \nonumber \\
    &=& \sum_j\!\int\!\!\text{d}\vec{r}\,'\,\chi^{\downarrow\uparrow}_{0,ij}(\vec{r}\,,\vec{r}\,';0)\,B_{\text{xc},j}(\vec{r}\,') \; ,
\end{eqnarray}
with the exchange-correlation splitting, $B_{\text{xc},i}(\vec{r}\,) = V_{\text{KS},i}^\uparrow(\vec{r}\,) - V_{\text{KS},i}^\downarrow(\vec{r}\,)$.
In the ASA the KS potential is spherical, so it is also consistent to take a spherical average of the magnetization,
\begin{eqnarray}
  m_i(r) &=& \!\int\!\!\text{d}\hat{r}\;m_i(\vec{r}\,) \nonumber\\
         &=& \!\int\!\!\text{d}\hat{r}\sum_j\!\int\!\!\text{d}\vec{r}\,'\,\chi^{\uparrow\downarrow}_{0,ij}(\vec{r}\,,\vec{r}\,';0)\,B_{\text{xc},j}(r') \nonumber\\
         &=& \!\int\!\!\text{d}\hat{r}\sum_j\!\int\!\!\text{d}\vec{r}\,'\,\chi^{\downarrow\uparrow}_{0,ij}(\vec{r}\,,\vec{r}\,';0)\,B_{\text{xc},j}(r') \; ,
\end{eqnarray}
and introducing the projection the spherical average of the $d$ magnetization turns out to be
\begin{eqnarray}
  m_d(r) &=& R_{d}^\uparrow(r;E_\text{F})\,R_{d}^\downarrow(r;E_\text{F}) \sum_{m_1m_2}\chi^{\uparrow\downarrow}_{0d,m_1m_2m_2m_1}(0) \times \nonumber\\
         & & \times \!\int\!\!\text{d}r'\,(r')^2\,R_{d}^\uparrow(r';E_\text{F})\,R_{d}^\downarrow(r';E_\text{F})\,B_{\text{xc}}(r') \nonumber\\
         &=& R_{d}^\uparrow(r;E_\text{F})\,R_{d}^\downarrow(r;E_\text{F})\,\bar{m}_d \; ,
\end{eqnarray}
using the orthogonality of the spherical harmonics.
This suggests the introduction of the spherical average of the KS susceptibility,
\begin{equation}
  \bar{\chi}^{\sigma\bar\sigma}_{0d}(\omega) = \sum_{m_1m_2}\chi^{\sigma\bar\sigma}_{0d,m_1m_2m_2m_1}(\omega) \; .
\end{equation}

In time-dependent density functional theory (TDDFT), the transverse magnetic susceptibility obeys the Dyson equation,
\begin{eqnarray}
  \chi^{\sigma\bar{\sigma}}_{ij}(\vec{r}\,,\vec{r}\,';\omega)
    &=& \chi^{\sigma\bar{\sigma}}_{0,ij}(\vec{r}\,,\vec{r}\,';\omega)
        + \sum_{pq}\!\int\!\!\text{d}\vec{r}_1\!\int\!\!\text{d}\vec{r}_2\; \times \nonumber\\
    & & \hspace{-6em} \times \chi^{\sigma\bar{\sigma}}_{0,ip}(\vec{r}\,,\vec{r}_1;\omega)\,U_{\text{xc},pq}(\vec{r}_1,\vec{r}_2;\omega)\,\chi^{\sigma\bar{\sigma}}_{qj}(\vec{r}_2,\vec{r}\,';\omega) \; ,
\end{eqnarray}
and in the adiabatic local density approximation the transverse xc kernel is simply given by
\begin{eqnarray}
  U_{\text{xc},ij}(\vec{r}\,,\vec{r}\,';\omega)
    &=& U_{\text{xc},i}(\vec{r}\,)\,\delta_{ij}\,\delta(\vec{r} - \vec{r}\,') \nonumber\\
    &=& \frac{B_{\text{xc},i}(\vec{r}\,)}{m_i(\vec{r}\,)}\,\delta_{ij}\,\delta(\vec{r} - \vec{r}\,') \; .
\end{eqnarray}
Returning to the ASA and the projection on the $d$ orbitals,
\begin{eqnarray}
  B_{\text{xc},i}(r) &=& \!\int\!\!\text{d}\hat{r}\;B_{\text{xc},i}(\vec{r}\,) = \!\int\!\!\text{d}\hat{r}\;U_{\text{xc},i}(\vec{r}\,)\,m_i(\vec{r}\,) \nonumber\\
                     &\approx& U_{\text{xc},d}(r)\,m_d(r) \nonumber\\
                     &=& U_{\text{xc},d}(r)\,R_{d}^\uparrow(r;E_\text{F})\,R_{d}^\downarrow(r;E_\text{F})\,\bar{m}_d \; .
\end{eqnarray}
From the magnetization sum rule we arrive at an effective one-parameter xc kernel,
\begin{eqnarray}
  \bar{U}_{\text{xc},d}
    &=& \!\int\!\!\text{d}r'\,(r')^2\,R_{d}^\uparrow(r';E_\text{F})\,R_{d}^\downarrow(r';E_\text{F})\,U_{\text{xc},d}(r') \times \nonumber\\
    & & \times R_{d}^\uparrow(r';E_\text{F})\,R_{d}^\downarrow(r';E_\text{F}) \nonumber\\
    &=& \Big(\bar{\chi}^{\uparrow\downarrow}_{0d}(0)\Big)^{-1} = \Big(\bar{\chi}^{\downarrow\uparrow}_{0d}(0)\Big)^{-1} \; ,
\end{eqnarray}
and the last two equalities must follow for consistency, which in practice define the kernel once the static KS susceptibility is known.

The transverse magnetic susceptibility is then represented as
\begin{equation}
  \bar{\chi}^{\sigma\bar{\sigma}}_{d}(\omega) = \Big(\bar{\chi}^{\sigma\bar\sigma}_{0d}(\omega)^{-1} - \bar{U}_{\text{xc},d}\Big)^{-1} \; .
\end{equation}

Let us turn our attention to the Dyson equation for the GF, including the self-energy describing the coupling to the magnetic excitations:
\begin{eqnarray}
  G_{ij}^\sigma(\vec{r}\,,\vec{r}\,';E)
    &=& G_{0,ij}^\sigma(\vec{r}\,,\vec{r}\,';E) + \sum_{pq}\!\int\!\!\text{d}\vec{r}_1\!\int\!\!\text{d}\vec{r}_2\; \times \nonumber\\
    & & \hspace{-6em}\times G_{0,ip}^\sigma(\vec{r}\,,\vec{r}_1;E)\,\Sigma_{pq}^\sigma(\vec{r}_1,\vec{r}_2;E)\,G_{qj}^\sigma(\vec{r}_2,\vec{r}\,';E) \; .
\end{eqnarray}
This will lead to the following matrix element, once the projection on the $d$ orbitals is introduced,

\begin{widetext}

\begin{equation}
  \Sigma_{d,mm'}^\sigma(E) = \!\int\!\!\text{d}\vec{r}\!\int\!\!\text{d}\vec{r}\,'\;R_{d}^\sigma(r;E_\text{F})\,Y_{2m}(\hat{r})\,\Sigma^\sigma(\vec{r}\,,\vec{r}\,';E)\,R_{d}^\sigma(r';E_\text{F})\,Y_{2m'}(\hat{r}') \; .
\end{equation}
The self-energy requires matrix elements of the following form, which simplify after replacing $G_0$ with its projected form, $U_{\text{xc}}$ and the susceptibility with their spherical averages:
\begin{align}
 &\!\int\!\!\text{d}\vec{r}\!\int\!\!\text{d}\vec{r}\,'\;R_{d}^\sigma(r;E_\text{F})\,Y_{2m}(\hat{r})\,U_{\text{xc}}(\vec{r}\,)\,G_{0}^{\bar\sigma}(\vec{r}\,,\vec{r}\,';E)\,\chi^{\bar\sigma\sigma}(\vec{r}\,,\vec{r}\,';E')\,U_{\text{xc}}(\vec{r}\,')\,R_{d}^\sigma(r';E_\text{F})\,Y_{2m'}(\hat{r}') \nonumber\\
&= \sum_{m_1m_2}\!\int\!\!\text{d}\vec{r}\!\int\!\!\text{d}\vec{r}\,'\;R_{d}^\sigma(r;E_\text{F})\,Y_{2m}(\hat{r})\,U_{\text{xc}}(r)
\,R_{d}^{\bar\sigma}(r;E_\text{F})\,Y_{2m_1}(\hat{r})\,G_{d,m_1m_2}^{\bar\sigma}(E)\,R_{d}^{\bar\sigma}(r';E_\text{F})\,Y_{2m_2}(\hat{r}') \times \nonumber\\
&\hspace{7em} \times R_{d}^\sigma(r;E_\text{F})\,R_{d}^{\bar\sigma}(r;E_\text{F})\,\bar{\chi}_d^{\bar\sigma\sigma}(E')\,R_{d}^\sigma(r';E_\text{F})\,R_{d}^{\bar\sigma}(r';E_\text{F})
\,U_{\text{xc}}(r')\,R_{d}^\sigma(r';E_\text{F})\,Y_{2m'}(\hat{r}') \phantom{\Big[}\nonumber\\
&= \bar{U}_{\text{xc},d}\,G_{d,mm'}^{\bar\sigma}(E)\,\bar{\chi}_d^{\bar\sigma\sigma}(E')\,\bar{U}_{\text{xc},d} \; ,
\end{align}

\end{widetext}

which is the form of the matrix elements of the self-energy quoted in Eq.~(1) in the main text.

\section{Density of states and self-energies for Cr, Mn, Fe, and Co adatoms on Cu(111)}

\begin{figure*}[ht!]
  \includegraphics[trim = 0mm 88mm 0mm 0mm, clip, width=0.9\textwidth]{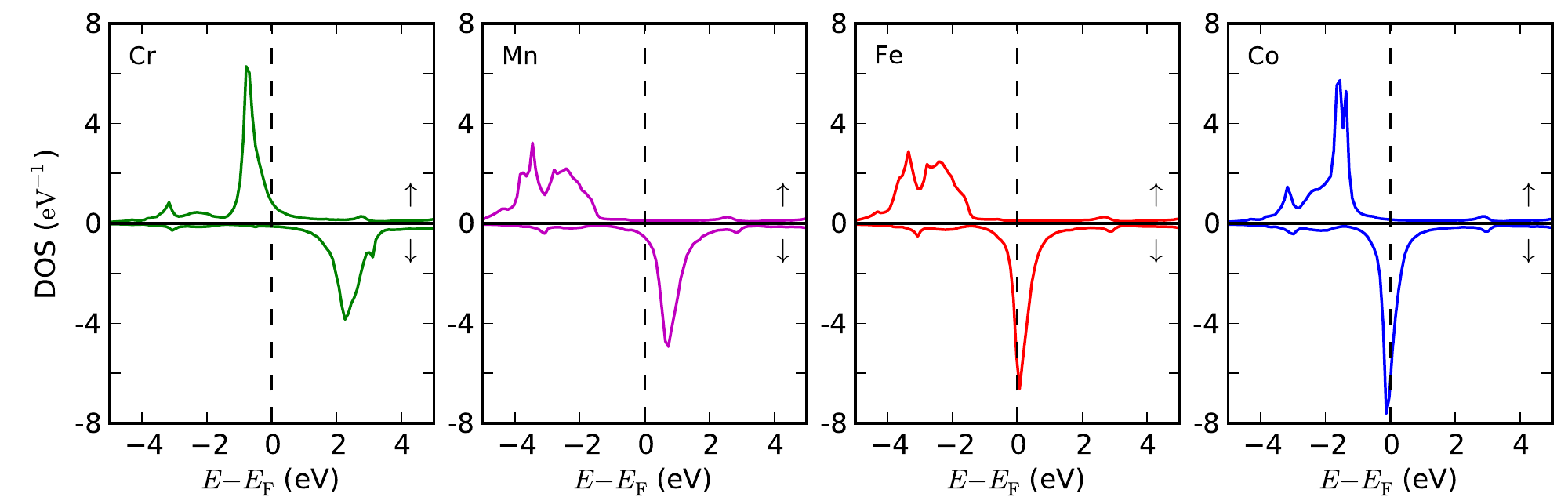}
  \caption{(Color online) The spin-resolved total density of states (DOS) are shown for the four adatoms (Cr, Mn, Fe, and Co, from left to right). The step height obtained for the imaginary part of the self-energy for a given spin channel 
is mainly determined by the local DOS of the opposite spin channel at the Fermi energy (dashed line).}
  \label{fig:tm_dos}
\end{figure*}

\begin{figure*}[tbp]
  \includegraphics[width=0.45\textwidth]{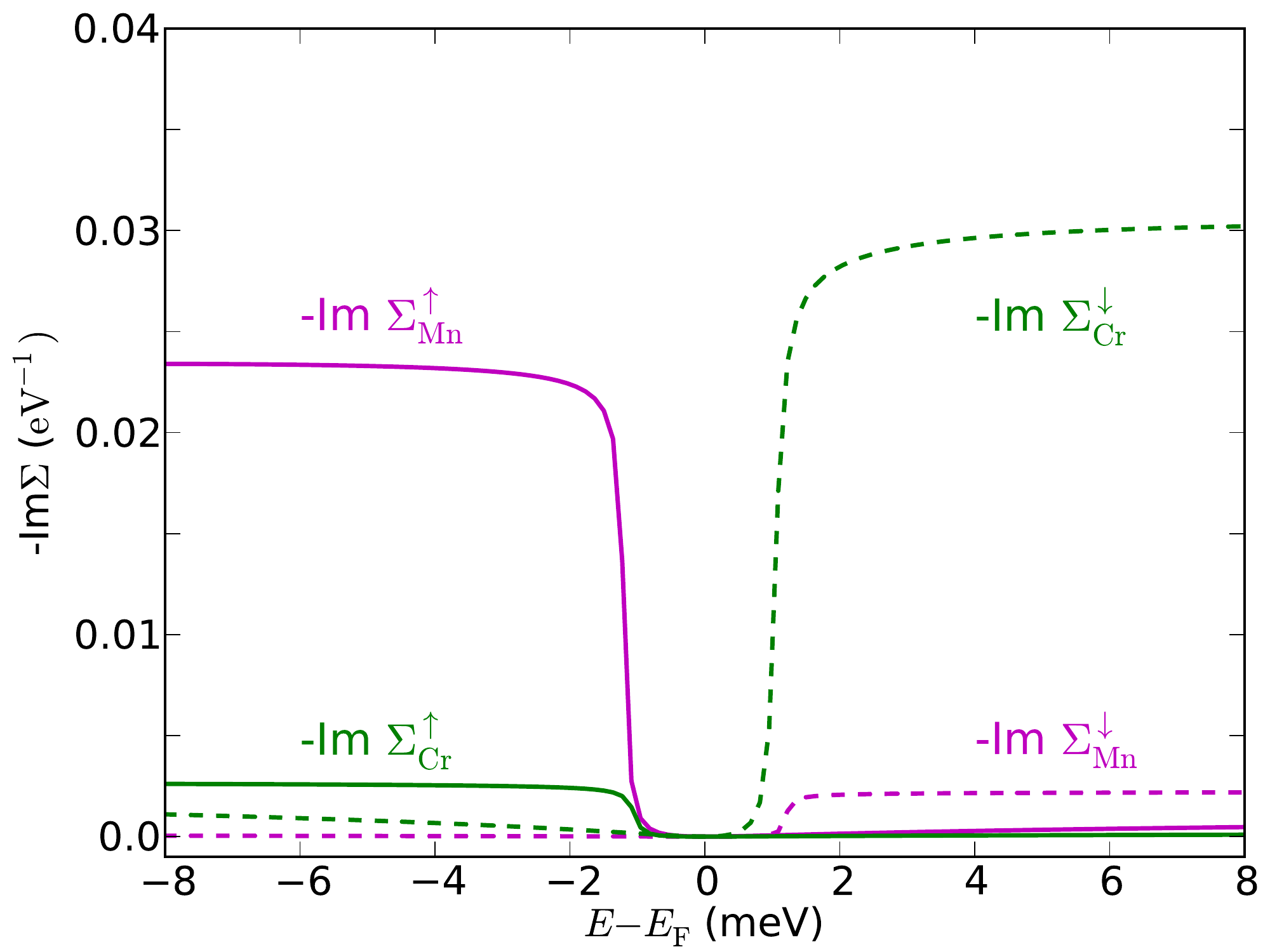}
  \includegraphics[width=0.45\textwidth]{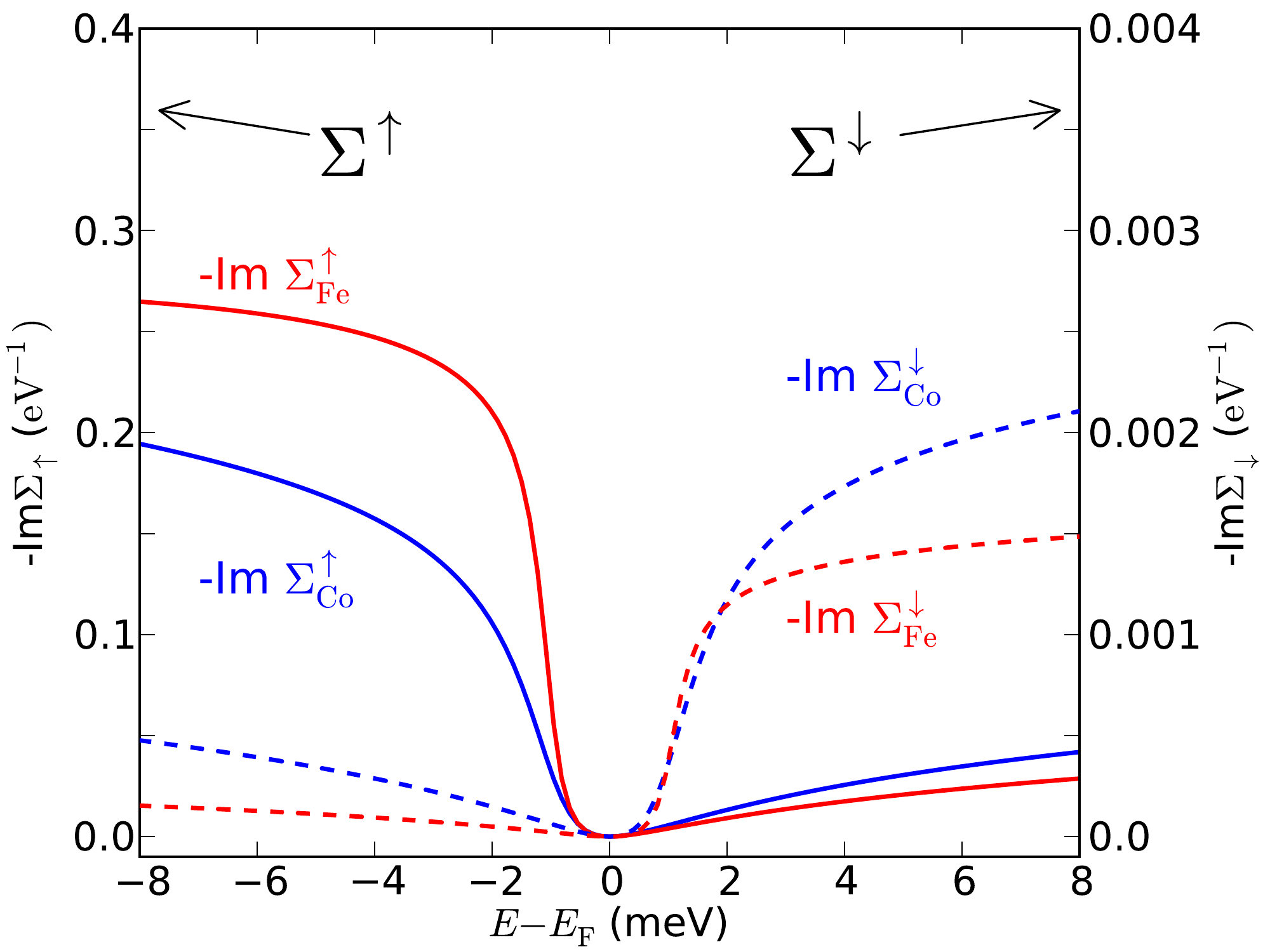}
  \caption{(Color online) The self-energies for the four investigated systems are shown. Whereas for Cr and Mn adatoms the self-energy step heights for the two spin channels are of the same order; they differ by a factor of 100 for the other two cases.}
  \label{fig:selfe_TM}
\end{figure*}

Following Eq.~(3) in the main text, characteristic features regarding the steps of the obtained imaginary parts of the self-energies, $\mathrm{Im}\Sigma$, can already be concluded from a brief analysis of the spin-resolved density of states (DOS), 
$n^\uparrow(E)$ and $n^\downarrow(E)$ for spin-up and spin-down, respectively. The step heights of the imaginary part of the self-energy for a given spin channel are weighted by the density of states of the opposite spin channel 
near the Fermi energy. In Fig.~\ref{fig:tm_dos} the spin-resolved DOS for Cr, Mn, Fe, and Co adatoms are shown. The majority-spin states are almost fully occupied. 
The minority-spin resonance shifts down in energy when increasing the $d$-electron occupation. Thus, for Fe and Co this resonance is located very close to the Fermi energy while for Cr and Mn the resonance is located much further above 
the Fermi energy. This explains the spin asymmetry observed in the step height of $\mathrm{Im}\Sigma$ as shown in Fig.~\ref{fig:selfe_TM}, where for the sake of comparison 
the auxiliary external magnetic field $B_0$ was kept the same for all four adatoms. Contrary to Mn, Fe, and Co adatoms, Cr adatom is the only case where $n_\mathrm{}^\uparrow(E_\mathrm{F}) > n_\mathrm{}^\downarrow(E_\mathrm{F})$ 
leading to $\mathrm{Im}\Sigma^\uparrow(E_\mathrm{F}) < \mathrm{Im}\Sigma^\downarrow(E_\mathrm{F})$. Since the spin asymmetry is large for the DOS of Fe and Co adatoms, the spin-dependent step heights of $\mathrm{Im}\Sigma$ differ by two orders of magnitude.

\section{Orbital-resolved analysis of the DOS for Fe adatoms}

\begin{figure*}[ht!]
  \includegraphics[trim = 0mm 88mm 0mm 0mm, clip, width=0.9\textwidth]{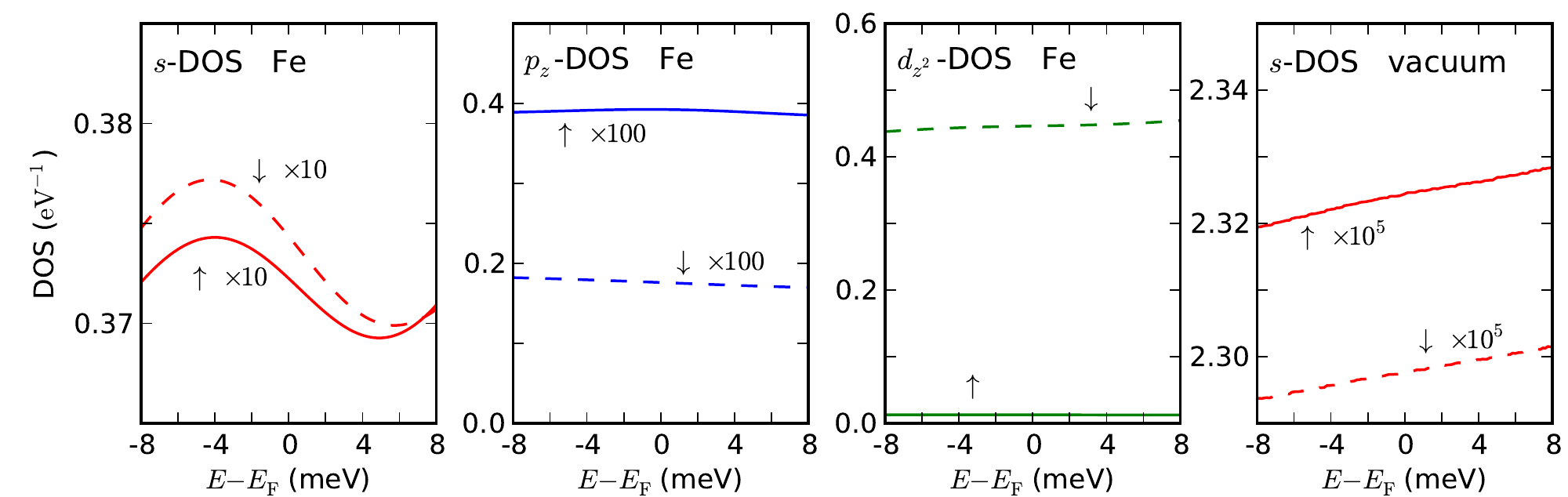}
  \caption{(Color online) The orbital-resolved DOS for Fe adatoms on the Cu(111) surface are shown ($s$, $p_z$, and $d_{z^2}$ orbitals) and compared to the DOS of the vacuum site \sunit{6.3}{\Angs} above the impurity ($s$ orbitals). 
The solid and dashed lines refer to spin up and spin down DOS, respectively. Only for the $p_z$ orbitals of the adatom the DOS for majority spin is larger than the one for the minority spin, which matches the weighting obtained for the vacuum site.}
  \label{fig:orbital_d_dos}
\end{figure*}

In Fig.~\ref{fig:orbital_d_dos} the orbital-resolved DOS is shown. Only states extending farthest into vacuum above the adatom are displayed ($s$, $p_z$, and $d_{z^2}$). 
Contrary to the $s$- and $d_{z^2}$-resolved DOS, for the $p_z$ states the majority-spin contribution is larger than the minority-spin contribution. 
This spin asymmetry in the magnitude of the DOS seem to be maintained in vacuum.  Whereas the $d_{z^2}$ state is dominant at the adatom it decays fast into vacuum due to its more localized character than the $p_z$ state. 
This is even more remarkable since the latter orbitals do only show a difference in the spin-resolved terms by a factor of 2 and are by more than two orders of magnitude smaller than those of the $d_{z^2}$ orbitals at the adatom. 
Such an observation does not necessarily remain true for all systems but shows that the tip position plays an important role when investigating the spin asymmetry of the inelastic spectra.




\section{What does a simple model predict for $\ud I/\ud V$ and $\ud^2 I/\ud V^2$?}

\begin{figure*}[ht!]
  \includegraphics[width=0.45\textwidth]{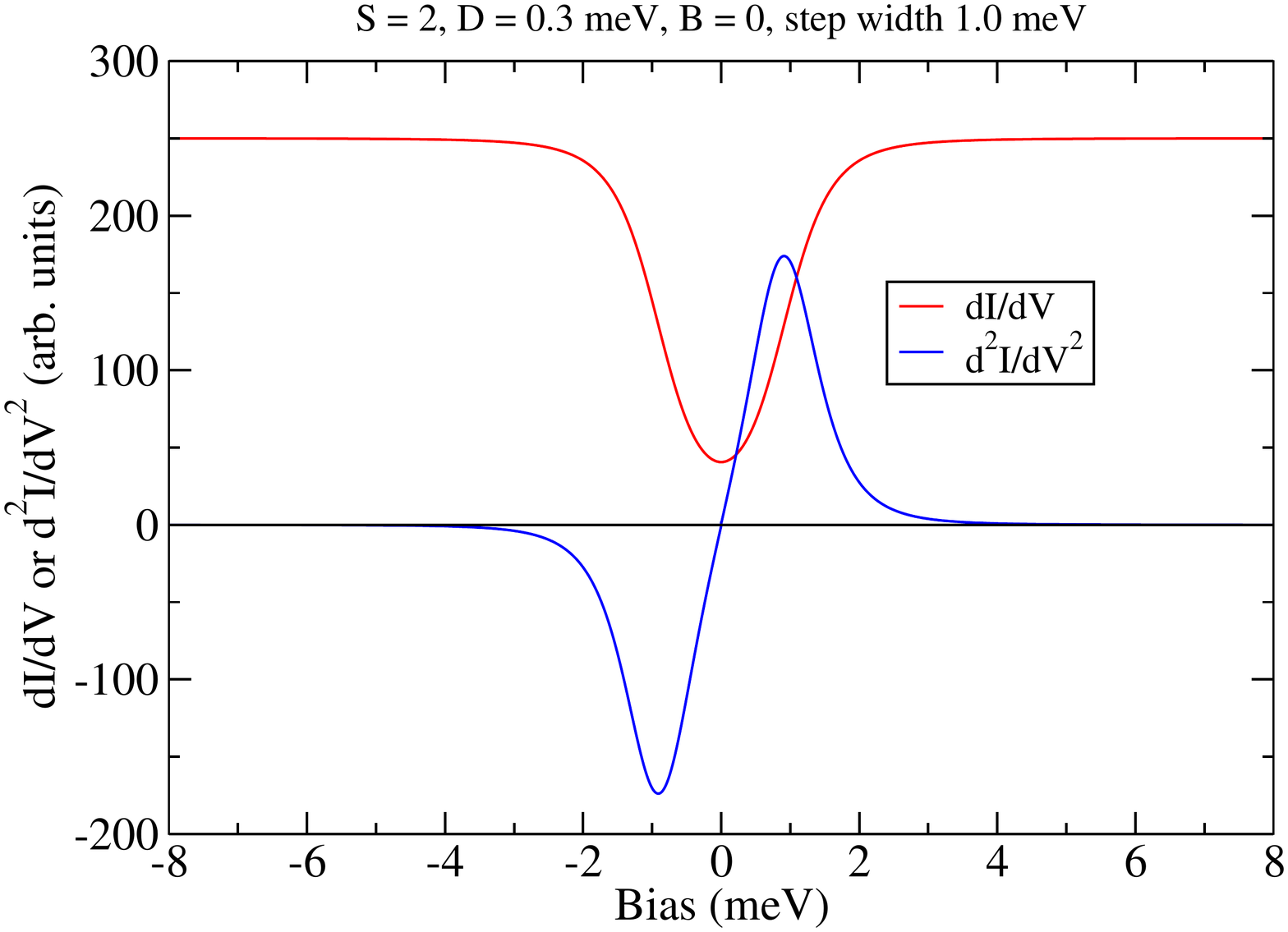}
  \includegraphics[width=0.45\textwidth]{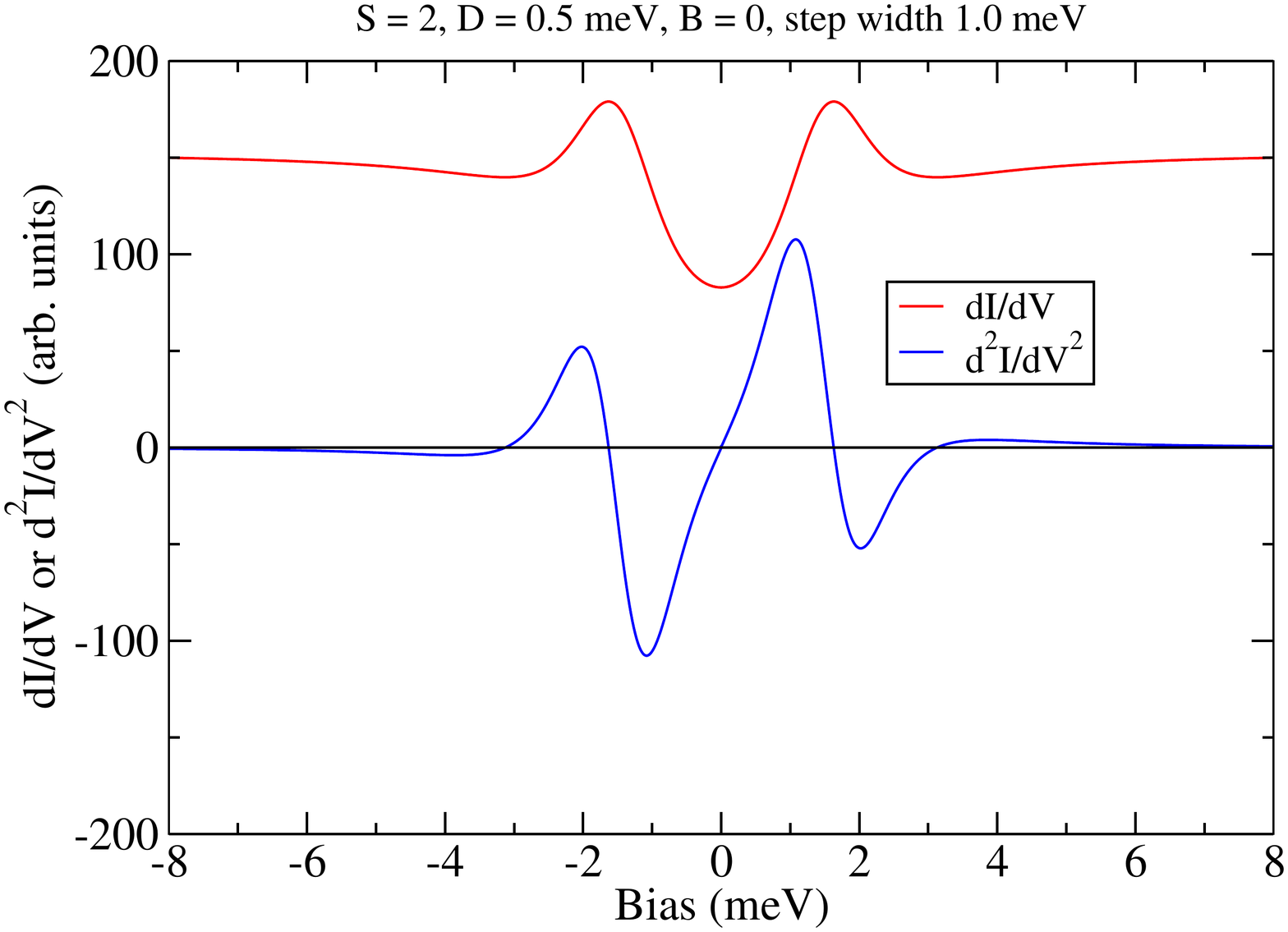}
  \caption{(Color online) Model forms for $\ud I/\ud V$ and $\ud^2 I/\ud V^2$. 
Left: a step in $\ud I/\ud V$ leads to a peak for $V > 0$ and a dip for $V < 0$.
Multiple steps lead to multiple peaks for $V > 0$ and multiple dips for $V < 0$, but never a dip for $V > 0$ or a peak for $V < 0$.
Right: a step with a shoulder or bump leads to a peak and dip pair at $V > 0$ and a dip and peak pair at $V < 0$.
Features always occur in $\pm V$ pairs; a peak for $V > 0$ must be accompanied by a dip for $V < 0$, in this picture.}
  \label{fig:step_model}
\end{figure*}

The simplest model able to predict steps in $\ud I/\ud V$ caused by inelastic tunneling via a magnetic adatom is that of a quantum spin coupled by exchange to the tunneling electrons (see, \textit{e.g.}, Refs.~[\onlinecite{LorenteGauyacqPRL09}], [\onlinecite{Fernandez-RossierPRL09}], and [\onlinecite{FranssonNL09}]). 
For the Fe adatom on the Cu(111) surface, the appropriate Hamiltonian is $\hat{\HH} = D\,\hat{S}_z^2 + B\,\hat{S}_z$, where $\hat{S}_z$ is the quantum angular momentum operator associated with states $|S\,M\rangle$ such that $\hat{S}^2 |S\,M\rangle = S(S+1) |S\,M\rangle$ and $\hat{S}_z |S\,M\rangle = M |S\,M\rangle$.
$D < 0$ describes the out-of-plane anisotropy easy axis found experimentally and theoretically by DFT calculations, and $B$ is the applied magnetic field in energy units. 
$S$ is usually chosen to be close to the computed spin magnetic moment from DFT; $S = 3/2$ or $S = 2$ are the values bracketing the calculated result ($1.85$).
The eigenvalues are then simply $E_M = D\,M^2 + B\,M$; inelastic transitions between eigenstates (caused by the tunneling electrons) obey the $M' = M \pm 1$ selection rules.
At very low temperature, the quantum spin is in its ground state ($B = 0$), either $|S\,+\!S\rangle$ or $|S\,-\!S\rangle$, or a superposition of the two.
The allowed transitions are then $|S\,+\!S\rangle \rightarrow |S\,+\!S\!-\!1\rangle$ and $|S\,-\!S\rangle \rightarrow |S\,-\!S\!+\!1\rangle$.
The energy difference $E_{S-1} - E_{S}$ corresponds to the threshold bias for inelastic transitions, marking the position of the steps in $\ud I/\ud V$.
The shape and width of the steps can only be given within this model by temperature broadening, which is too small (typical experimental temperatures are $\sim 1$~K $\sim 0.1$ meV).
As far as this model goes, other step shapes or broadening mechanisms are not taken into account.
Our TD-DFT calculations provide an alternative and realistic route to the step width and shape, via interaction between itinerant electrons and spin excitations, contained in the self-energy.
To interpret the experimental data in Fig.~\ref{fig:results_Susc_Selfe_DOS}(e) of the main text, we present two generic step shapes commonly seen in experiment (and in our calculations; see Fig.~\ref{fig:DOS_TM_vac} in the main text) in Fig.~\ref{fig:step_model} above. 
Note that the artificial broadening used in generating these figures is $\sim 1$ meV, comparable with the experimental and \textit{ab initio} line widths. 
The threshold bias is assumed to be given by $\pm | E_{S-1} - E_{S} |$ in the discussed model.
The step shapes are meant for illustration purposes only.
A broadened step would lead to a peak for $V > 0$ and a dip for $V < 0$ (Fig.~\ref{fig:step_model}, left).
A broadened step topped with a bump would lead to a peak and dip pair for $V > 0$ and a dip and peak pair for $V < 0$ (Fig.~\ref{fig:step_model}, right).
This matches the peak and dip, and dip and peak pairs seen in the experimental data.
The lone peak around $+5$~meV in the experimental data, though, would require a matching dip around $-5$~meV, according to this model.
This one-sided behavior in the conductance is readily explained by a satellite arising from the self-energy, as detailed in the main text.

\section{Magnetized tip -- a simple approach}

The Tersoff-Hamann approximation~\cite{TersoffHamannPRL83} relates the conductance to the density of states from the tip as well as from the probed adatom,
\begin{eqnarray}
  \frac{\mathrm{d}I}{\mathrm{d}V} \propto &&\Big[ n_\mathrm{tip}^\uparrow \cdot n_\mathrm{adatom}^\uparrow(E_\mathrm{F}+V) \nonumber\\
  &&+ n_\mathrm{tip}^\downarrow \cdot n_\mathrm{adatom}^\downarrow(E_\mathrm{F}+V) \Big] \; .
\end{eqnarray}
 For a nonmagnetic tip, one has
\begin{eqnarray}
  n_\mathrm{tip}^\uparrow = n_\mathrm{tip}^\downarrow = \frac{N}{2} \; ,
\end{eqnarray}
with $N$ the total density of states of the tip. This leads to
\begin{eqnarray}
  \frac{\mathrm{d}I}{\mathrm{d}V} \propto \left[ n_\mathrm{adatom}^\uparrow + n_\mathrm{adatom}^\downarrow \right] \; .
\end{eqnarray}

For a magnetic tip one finds a nonvanishing polarization:
\begin{eqnarray}
  P = \frac{n_\mathrm{tip}^\uparrow - n_\mathrm{tip}^\downarrow}{N} \; .
\end{eqnarray}
Thus, we have
\begin{eqnarray}
  n_\mathrm{tip}^\uparrow   &=& \frac{N}{2} (1+P) \; , \\
  n_\mathrm{tip}^\downarrow &=& \frac{N}{2} (1-P) \; ,
\end{eqnarray}
and depending on the sign of $P$, one spin channel gives a larger contribution to the spectrum than the other,
\begin{eqnarray}
  \frac{\mathrm{d}I}{\mathrm{d}V} \propto \left[ (1+P) \cdot n_\mathrm{adatom}^\uparrow + (1-P) \cdot n_\mathrm{adatom}^\downarrow \right] \; .
\end{eqnarray}
For the figure shown in the main text, Fig.~\ref{fig:results_Susc_Selfe_DOS}(e), we set $P=-50$ \%, meaning that $n_\mathrm{adatom}^\downarrow$ has three times more weight than $n_\mathrm{adatom}^\uparrow$.

\end{appendix}

\bibliography{bibliography}

\begin{thebibliography}{52}%
\makeatletter
\providecommand \@ifxundefined [1]{%
 \@ifx{#1\undefined}
}%
\providecommand \@ifnum [1]{%
 \ifnum #1\expandafter \@firstoftwo
 \else \expandafter \@secondoftwo
 \fi
}%
\providecommand \@ifx [1]{%
 \ifx #1\expandafter \@firstoftwo
 \else \expandafter \@secondoftwo
 \fi
}%
\providecommand \natexlab [1]{#1}%
\providecommand \enquote  [1]{``#1''}%
\providecommand \bibnamefont  [1]{#1}%
\providecommand \bibfnamefont [1]{#1}%
\providecommand \citenamefont [1]{#1}%
\providecommand \href@noop [0]{\@secondoftwo}%
\providecommand \href [0]{\begingroup \@sanitize@url \@href}%
\providecommand \@href[1]{\@@startlink{#1}\@@href}%
\providecommand \@@href[1]{\endgroup#1\@@endlink}%
\providecommand \@sanitize@url [0]{\catcode `\\12\catcode `\$12\catcode
  `\&12\catcode `\#12\catcode `\^12\catcode `\_12\catcode `\%12\relax}%
\providecommand \@@startlink[1]{}%
\providecommand \@@endlink[0]{}%
\providecommand \url  [0]{\begingroup\@sanitize@url \@url }%
\providecommand \@url [1]{\endgroup\@href {#1}{\urlprefix }}%
\providecommand \urlprefix  [0]{URL }%
\providecommand \Eprint [0]{\href }%
\@ifxundefined \urlstyle {%
  \providecommand \doi  [0]{\begingroup \@sanitize@url \@doi}%
  \providecommand \@doi [1]{\endgroup \@@startlink {\doibase
  #1}doi:\discretionary {}{}{}#1\@@endlink }%
}{%
  \providecommand \doi  [0]{doi:\discretionary{}{}{}\begingroup
  \urlstyle{rm}\Url }%
}%
\providecommand \doibase [0]{http://dx.doi.org/}%
\providecommand \Doi [0]{\begingroup \@sanitize@url \@Doi }%
\providecommand \@Doi  [1]{\endgroup\@@startlink{\doibase#1}\@@Doi}%
\providecommand \@@Doi [1]{#1\@@endlink}%
\providecommand \selectlanguage [0]{\@gobble}%
\providecommand \bibinfo  [0]{\@secondoftwo}%
\providecommand \bibfield  [0]{\@secondoftwo}%
\providecommand \translation [1]{[#1]}%
\providecommand \BibitemOpen [0]{}%
\providecommand \bibitemStop [0]{}%
\providecommand \bibitemNoStop [0]{.\EOS\space}%
\providecommand \EOS [0]{\spacefactor3000\relax}%
\providecommand \BibitemShut  [1]{\csname bibitem#1\endcsname}%
\bibitem [{\citenamefont {Zutic}\ \emph {et~al.}(2004)\citenamefont {Zutic},
  \citenamefont {Fabian},\ and\ \citenamefont {Sarma}}]{Fabian+RMP04}%
  \BibitemOpen
  \bibfield  {author} {\bibinfo {author} {\bibfnamefont {I.}~\bibnamefont
  {Zutic}}, \bibinfo {author} {\bibfnamefont {J.}~\bibnamefont {Fabian}}, \
  and\ \bibinfo {author} {\bibfnamefont {S.~D.}\ \bibnamefont {Sarma}},\
  }\href@noop {} {\bibfield  {journal} {\bibinfo  {journal} {Rev. Mod. Phys.},\
  }\textbf {\bibinfo {volume} {76}},\ \bibinfo {pages} {323} (\bibinfo {year}
  {2004})}\BibitemShut {NoStop}%
\bibitem [{\citenamefont {Moriya}(1985)}]{MoriyaSpringerSeries85}%
  \BibitemOpen
  \bibfield  {author} {\bibinfo {author} {\bibfnamefont {T.}~\bibnamefont
  {Moriya}},\ }\href@noop {} {\bibfield  {journal} {\bibinfo  {journal} {{\it
  Spin Fluctuations in Itinerant Electron Magnetism}, Springer-Verlag}}
  (\bibinfo {year} {1985})}\BibitemShut {NoStop}%
\bibitem [{\citenamefont {Mazin}\ \emph {et~al.}(2008)\citenamefont {Mazin},
  \citenamefont {Singh}, \citenamefont {Johannes},\ and\ \citenamefont
  {Du}}]{Mazin+PRL08}%
  \BibitemOpen
  \bibfield  {author} {\bibinfo {author} {\bibfnamefont {I.~I.}\ \bibnamefont
  {Mazin}}, \bibinfo {author} {\bibfnamefont {D.~J.}\ \bibnamefont {Singh}},
  \bibinfo {author} {\bibfnamefont {M.~D.}\ \bibnamefont {Johannes}}, \ and\
  \bibinfo {author} {\bibfnamefont {M.~H.}\ \bibnamefont {Du}},\ }\href@noop {}
  {\bibfield  {journal} {\bibinfo  {journal} {Phys. Rev. Lett.},\ }\textbf
  {\bibinfo {volume} {101}},\ \bibinfo {pages} {057003} (\bibinfo {year}
  {2008})}\BibitemShut {NoStop}%
\bibitem [{\citenamefont {Degatto}(1994)}]{DagattoRMP94}%
  \BibitemOpen
  \bibfield  {author} {\bibinfo {author} {\bibfnamefont {E.}~\bibnamefont
  {Degatto}},\ }\href@noop {} {\bibfield  {journal} {\bibinfo  {journal} {Rev.
  Mod. Phys.},\ }\textbf {\bibinfo {volume} {66}},\ \bibinfo {pages} {763}
  (\bibinfo {year} {1994})}\BibitemShut {NoStop}%
\bibitem [{\citenamefont {Khajetoorians}\ \emph
  {et~al.}(2013){\natexlab{a}}\citenamefont {Khajetoorians}, \citenamefont
  {Baxevanis}, \citenamefont {H\"ubner}, \citenamefont {Schlenk}, \citenamefont
  {Krause}, \citenamefont {Wehling}, \citenamefont {Lounis}, \citenamefont
  {Lichtenstein}, \citenamefont {Pfannkuche}, \citenamefont {Wiebe},\ and\
  \citenamefont {Wiesendanger}}]{Khajetoorians+Science13}%
  \BibitemOpen
  \bibfield  {author} {\bibinfo {author} {\bibfnamefont {A.~A.}\ \bibnamefont
  {Khajetoorians}}, \bibinfo {author} {\bibfnamefont {B.}~\bibnamefont
  {Baxevanis}}, \bibinfo {author} {\bibfnamefont {C.}~\bibnamefont {H\"ubner}},
  \bibinfo {author} {\bibfnamefont {T.}~\bibnamefont {Schlenk}}, \bibinfo
  {author} {\bibfnamefont {S.}~\bibnamefont {Krause}}, \bibinfo {author}
  {\bibfnamefont {T.~O.}\ \bibnamefont {Wehling}}, \bibinfo {author}
  {\bibfnamefont {S.}~\bibnamefont {Lounis}}, \bibinfo {author} {\bibfnamefont
  {A.}~\bibnamefont {Lichtenstein}}, \bibinfo {author} {\bibfnamefont
  {D.}~\bibnamefont {Pfannkuche}}, \bibinfo {author} {\bibfnamefont
  {J.}~\bibnamefont {Wiebe}}, \ and\ \bibinfo {author} {\bibfnamefont
  {R.}~\bibnamefont {Wiesendanger}},\ }\href@noop {} {\bibfield  {journal}
  {\bibinfo  {journal} {Science},\ }\textbf {\bibinfo {volume} {339}},\
  \bibinfo {pages} {55} (\bibinfo {year} {2013}{\natexlab{a}})}\BibitemShut
  {NoStop}%
\bibitem [{\citenamefont {Sch\"afer}\ \emph {et~al.}(2004)\citenamefont
  {Sch\"afer}, \citenamefont {Schrupp}, \citenamefont {Rotenberg},
  \citenamefont {Rossnagel}, \citenamefont {Koh}, \citenamefont {Blaha},\ and\
  \citenamefont {Claessen}}]{Schaefer+PRL04}%
  \BibitemOpen
  \bibfield  {author} {\bibinfo {author} {\bibfnamefont {J.}~\bibnamefont
  {Sch\"afer}}, \bibinfo {author} {\bibfnamefont {D.}~\bibnamefont {Schrupp}},
  \bibinfo {author} {\bibfnamefont {E.}~\bibnamefont {Rotenberg}}, \bibinfo
  {author} {\bibfnamefont {K.}~\bibnamefont {Rossnagel}}, \bibinfo {author}
  {\bibfnamefont {H.}~\bibnamefont {Koh}}, \bibinfo {author} {\bibfnamefont
  {P.}~\bibnamefont {Blaha}}, \ and\ \bibinfo {author} {\bibfnamefont
  {R.}~\bibnamefont {Claessen}},\ }\href@noop {} {\bibfield  {journal}
  {\bibinfo  {journal} {Phys. Rev. Lett.},\ }\textbf {\bibinfo {volume} {92}},\
  \bibinfo {pages} {097205} (\bibinfo {year} {2004})}\BibitemShut {NoStop}%
\bibitem [{\citenamefont {Cui}\ \emph {et~al.}(2007)\citenamefont {Cui},
  \citenamefont {Shimada}, \citenamefont {Hoesch}, \citenamefont {Sakisaka},
  \citenamefont {Kato}, \citenamefont {Aiura}, \citenamefont {Negishi},
  \citenamefont {Higashiguchi}, \citenamefont {Miura}, \citenamefont
  {Namatame},\ and\ \citenamefont {Taniguchi}}]{Cui+JMM07}%
  \BibitemOpen
  \bibfield  {author} {\bibinfo {author} {\bibfnamefont {X.~Y.}\ \bibnamefont
  {Cui}}, \bibinfo {author} {\bibfnamefont {K.}~\bibnamefont {Shimada}},
  \bibinfo {author} {\bibfnamefont {M.}~\bibnamefont {Hoesch}}, \bibinfo
  {author} {\bibfnamefont {Y.}~\bibnamefont {Sakisaka}}, \bibinfo {author}
  {\bibfnamefont {H.}~\bibnamefont {Kato}}, \bibinfo {author} {\bibfnamefont
  {Y.}~\bibnamefont {Aiura}}, \bibinfo {author} {\bibfnamefont
  {S.}~\bibnamefont {Negishi}}, \bibinfo {author} {\bibfnamefont
  {M.}~\bibnamefont {Higashiguchi}}, \bibinfo {author} {\bibfnamefont
  {Y.}~\bibnamefont {Miura}}, \bibinfo {author} {\bibfnamefont
  {H.}~\bibnamefont {Namatame}}, \ and\ \bibinfo {author} {\bibfnamefont
  {M.}~\bibnamefont {Taniguchi}},\ }\href@noop {} {\bibfield  {journal}
  {\bibinfo  {journal} {J. Mag. Mat.},\ }\textbf {\bibinfo {volume} {310}},\
  \bibinfo {pages} {1617} (\bibinfo {year} {2007})}\BibitemShut {NoStop}%
\bibitem [{\citenamefont {Hofmann}\ \emph {et~al.}(2009)\citenamefont
  {Hofmann}, \citenamefont {Cui}, \citenamefont {Sch\"afer}, \citenamefont
  {Meyer}, \citenamefont {H\"opfner}, \citenamefont {Blumenstein},
  \citenamefont {Paul}, \citenamefont {Patthey}, \citenamefont {Rotenberg},
  \citenamefont {B\"unemann}, \citenamefont {Gebhard}, \citenamefont {Ohm},
  \citenamefont {Weber},\ and\ \citenamefont {Claessen}}]{Hofmann+PRL09}%
  \BibitemOpen
  \bibfield  {author} {\bibinfo {author} {\bibfnamefont {A.}~\bibnamefont
  {Hofmann}}, \bibinfo {author} {\bibfnamefont {X.~Y.}\ \bibnamefont {Cui}},
  \bibinfo {author} {\bibfnamefont {J.}~\bibnamefont {Sch\"afer}}, \bibinfo
  {author} {\bibfnamefont {S.}~\bibnamefont {Meyer}}, \bibinfo {author}
  {\bibfnamefont {P.}~\bibnamefont {H\"opfner}}, \bibinfo {author}
  {\bibfnamefont {C.}~\bibnamefont {Blumenstein}}, \bibinfo {author}
  {\bibfnamefont {M.}~\bibnamefont {Paul}}, \bibinfo {author} {\bibfnamefont
  {L.}~\bibnamefont {Patthey}}, \bibinfo {author} {\bibfnamefont
  {E.}~\bibnamefont {Rotenberg}}, \bibinfo {author} {\bibfnamefont
  {J.}~\bibnamefont {B\"unemann}}, \bibinfo {author} {\bibfnamefont
  {F.}~\bibnamefont {Gebhard}}, \bibinfo {author} {\bibfnamefont
  {T.}~\bibnamefont {Ohm}}, \bibinfo {author} {\bibfnamefont {W.}~\bibnamefont
  {Weber}}, \ and\ \bibinfo {author} {\bibfnamefont {R.}~\bibnamefont
  {Claessen}},\ }\href@noop {} {\bibfield  {journal} {\bibinfo  {journal}
  {Phys. Rev. Lett.},\ }\textbf {\bibinfo {volume} {102}},\ \bibinfo {pages}
  {187204} (\bibinfo {year} {2009})}\BibitemShut {NoStop}%
\bibitem [{\citenamefont {Heinrich}\ \emph {et~al.}(2004)\citenamefont
  {Heinrich}, \citenamefont {Gupta}, \citenamefont {Lutz},\ and\ \citenamefont
  {Eigler}}]{Heinrich+Science04}%
  \BibitemOpen
  \bibfield  {author} {\bibinfo {author} {\bibfnamefont {A.~J.}\ \bibnamefont
  {Heinrich}}, \bibinfo {author} {\bibfnamefont {J.~A.}\ \bibnamefont {Gupta}},
  \bibinfo {author} {\bibfnamefont {C.~P.}\ \bibnamefont {Lutz}}, \ and\
  \bibinfo {author} {\bibfnamefont {D.~M.}\ \bibnamefont {Eigler}},\
  }\href@noop {} {\bibfield  {journal} {\bibinfo  {journal} {Science},\
  }\textbf {\bibinfo {volume} {306}},\ \bibinfo {pages} {466} (\bibinfo {year}
  {2004})}\BibitemShut {NoStop}%
\bibitem [{\citenamefont {Hirjibehedin}\ \emph {et~al.}(2006)\citenamefont
  {Hirjibehedin}, \citenamefont {Lutz},\ and\ \citenamefont
  {Heinrich}}]{Hirjibehedin+Science06}%
  \BibitemOpen
  \bibfield  {author} {\bibinfo {author} {\bibfnamefont {C.~F.}\ \bibnamefont
  {Hirjibehedin}}, \bibinfo {author} {\bibfnamefont {C.~P.}\ \bibnamefont
  {Lutz}}, \ and\ \bibinfo {author} {\bibfnamefont {A.~J.}\ \bibnamefont
  {Heinrich}},\ }\href@noop {} {\bibfield  {journal} {\bibinfo  {journal}
  {Science},\ }\textbf {\bibinfo {volume} {312}},\ \bibinfo {pages} {1021}
  (\bibinfo {year} {2006})}\BibitemShut {NoStop}%
\bibitem [{\citenamefont {Balashov}\ \emph {et~al.}(2009)\citenamefont
  {Balashov}, \citenamefont {Schuh}, \citenamefont {Tak\'acs}, \citenamefont
  {Ernst}, \citenamefont {Ostanin}, \citenamefont {Henk}, \citenamefont
  {Mertig}, \citenamefont {Bruno}, \citenamefont {Miyamachi}, \citenamefont
  {Suga},\ and\ \citenamefont {Wulfhekel}}]{Balashov+PRL09}%
  \BibitemOpen
  \bibfield  {author} {\bibinfo {author} {\bibfnamefont {T.}~\bibnamefont
  {Balashov}}, \bibinfo {author} {\bibfnamefont {T.}~\bibnamefont {Schuh}},
  \bibinfo {author} {\bibfnamefont {A.~F.}\ \bibnamefont {Tak\'acs}}, \bibinfo
  {author} {\bibfnamefont {A.}~\bibnamefont {Ernst}}, \bibinfo {author}
  {\bibfnamefont {S.}~\bibnamefont {Ostanin}}, \bibinfo {author} {\bibfnamefont
  {J.}~\bibnamefont {Henk}}, \bibinfo {author} {\bibfnamefont {I.}~\bibnamefont
  {Mertig}}, \bibinfo {author} {\bibfnamefont {P.}~\bibnamefont {Bruno}},
  \bibinfo {author} {\bibfnamefont {T.}~\bibnamefont {Miyamachi}}, \bibinfo
  {author} {\bibfnamefont {S.}~\bibnamefont {Suga}}, \ and\ \bibinfo {author}
  {\bibfnamefont {W.}~\bibnamefont {Wulfhekel}},\ }\href@noop {} {\bibfield
  {journal} {\bibinfo  {journal} {Phys. Rev. Lett.},\ }\textbf {\bibinfo
  {volume} {102}},\ \bibinfo {pages} {257203} (\bibinfo {year}
  {2009})}\BibitemShut {NoStop}%
\bibitem [{\citenamefont {Khajetoorians}\ \emph {et~al.}(2011)\citenamefont
  {Khajetoorians}, \citenamefont {Lounis}, \citenamefont {Chilian},
  \citenamefont {Costa}, \citenamefont {Zhou}, \citenamefont {Mills},
  \citenamefont {Wiebe},\ and\ \citenamefont
  {Wiesendanger}}]{Khajetoorians+PRL11}%
  \BibitemOpen
  \bibfield  {author} {\bibinfo {author} {\bibfnamefont {A.~A.}\ \bibnamefont
  {Khajetoorians}}, \bibinfo {author} {\bibfnamefont {S.}~\bibnamefont
  {Lounis}}, \bibinfo {author} {\bibfnamefont {B.}~\bibnamefont {Chilian}},
  \bibinfo {author} {\bibfnamefont {A.~T.}\ \bibnamefont {Costa}}, \bibinfo
  {author} {\bibfnamefont {L.}~\bibnamefont {Zhou}}, \bibinfo {author}
  {\bibfnamefont {D.~L.}\ \bibnamefont {Mills}}, \bibinfo {author}
  {\bibfnamefont {J.}~\bibnamefont {Wiebe}}, \ and\ \bibinfo {author}
  {\bibfnamefont {R.}~\bibnamefont {Wiesendanger}},\ }\href@noop {} {\bibfield
  {journal} {\bibinfo  {journal} {Phys. Rev. Lett.},\ }\textbf {\bibinfo
  {volume} {106}},\ \bibinfo {pages} {037205} (\bibinfo {year}
  {2011})}\BibitemShut {NoStop}%
\bibitem [{\citenamefont {Chilian}\ \emph {et~al.}(2011)\citenamefont
  {Chilian}, \citenamefont {Khajetoorians}, \citenamefont {Lounis},
  \citenamefont {Costa}, \citenamefont {Mills}, \citenamefont {Wiebe},\ and\
  \citenamefont {Wiesendanger}}]{Chilian+PRB11}%
  \BibitemOpen
  \bibfield  {author} {\bibinfo {author} {\bibfnamefont {B.}~\bibnamefont
  {Chilian}}, \bibinfo {author} {\bibfnamefont {A.~A.}\ \bibnamefont
  {Khajetoorians}}, \bibinfo {author} {\bibfnamefont {S.}~\bibnamefont
  {Lounis}}, \bibinfo {author} {\bibfnamefont {A.~T.}\ \bibnamefont {Costa}},
  \bibinfo {author} {\bibfnamefont {D.~L.}\ \bibnamefont {Mills}}, \bibinfo
  {author} {\bibfnamefont {J.}~\bibnamefont {Wiebe}}, \ and\ \bibinfo {author}
  {\bibfnamefont {R.}~\bibnamefont {Wiesendanger}},\ }\href@noop {} {\bibfield
  {journal} {\bibinfo  {journal} {Phys. Rev. B},\ }\textbf {\bibinfo {volume}
  {84}},\ \bibinfo {pages} {212401} (\bibinfo {year} {2011})}\BibitemShut
  {NoStop}%
\bibitem [{\citenamefont {Bryant}\ \emph {et~al.}(2013)\citenamefont {Bryant},
  \citenamefont {Spinelli}, \citenamefont {Wagenaar}, \citenamefont {Gerrits},\
  and\ \citenamefont {Otte}}]{Otte+PRL13}%
  \BibitemOpen
  \bibfield  {author} {\bibinfo {author} {\bibfnamefont {B.}~\bibnamefont
  {Bryant}}, \bibinfo {author} {\bibfnamefont {A.}~\bibnamefont {Spinelli}},
  \bibinfo {author} {\bibfnamefont {J.~J.~T.}\ \bibnamefont {Wagenaar}},
  \bibinfo {author} {\bibfnamefont {M.}~\bibnamefont {Gerrits}}, \ and\
  \bibinfo {author} {\bibfnamefont {A.~F.}\ \bibnamefont {Otte}},\ }\href@noop
  {} {\bibfield  {journal} {\bibinfo  {journal} {Phys. Rev. Lett.},\ }\textbf
  {\bibinfo {volume} {111}},\ \bibinfo {pages} {127203} (\bibinfo {year}
  {2013})}\BibitemShut {NoStop}%
\bibitem [{\citenamefont {Khajetoorians}\ \emph
  {et~al.}(2013){\natexlab{b}}\citenamefont {Khajetoorians}, \citenamefont
  {Schlenk}, \citenamefont {Schweflinghaus}, \citenamefont {{dos Santos Dias}},
  \citenamefont {Steinbrecher}, \citenamefont {Bouhassoune}, \citenamefont
  {Lounis}, \citenamefont {Wiebe},\ and\ \citenamefont
  {Wiesendanger}}]{Khajetoorians+PRL13}%
  \BibitemOpen
  \bibfield  {author} {\bibinfo {author} {\bibfnamefont {A.~A.}\ \bibnamefont
  {Khajetoorians}}, \bibinfo {author} {\bibfnamefont {T.}~\bibnamefont
  {Schlenk}}, \bibinfo {author} {\bibfnamefont {B.}~\bibnamefont
  {Schweflinghaus}}, \bibinfo {author} {\bibfnamefont {M.}~\bibnamefont {{dos
  Santos Dias}}}, \bibinfo {author} {\bibfnamefont {M.}~\bibnamefont
  {Steinbrecher}}, \bibinfo {author} {\bibfnamefont {M.}~\bibnamefont
  {Bouhassoune}}, \bibinfo {author} {\bibfnamefont {S.}~\bibnamefont {Lounis}},
  \bibinfo {author} {\bibfnamefont {J.}~\bibnamefont {Wiebe}}, \ and\ \bibinfo
  {author} {\bibfnamefont {R.}~\bibnamefont {Wiesendanger}},\ }\href@noop {}
  {\bibfield  {journal} {\bibinfo  {journal} {Phys. Rev. Lett.},\ }\textbf
  {\bibinfo {volume} {111}},\ \bibinfo {pages} {157204} (\bibinfo {year}
  {2013}{\natexlab{b}})}\BibitemShut {NoStop}%
\bibitem [{\citenamefont {Mills}\ and\ \citenamefont
  {Lederer}(1967)}]{MillsLedererPR67}%
  \BibitemOpen
  \bibfield  {author} {\bibinfo {author} {\bibfnamefont {D.~L.}\ \bibnamefont
  {Mills}}\ and\ \bibinfo {author} {\bibfnamefont {P.}~\bibnamefont
  {Lederer}},\ }\href@noop {} {\bibfield  {journal} {\bibinfo  {journal} {Phys.
  Rev.},\ }\textbf {\bibinfo {volume} {160}},\ \bibinfo {pages} {590} (\bibinfo
  {year} {1967})}\BibitemShut {NoStop}%
\bibitem [{\citenamefont {Muniz}\ and\ \citenamefont
  {Mills}(2003)}]{MunizMillsPRB03}%
  \BibitemOpen
  \bibfield  {author} {\bibinfo {author} {\bibfnamefont {R.~B.}\ \bibnamefont
  {Muniz}}\ and\ \bibinfo {author} {\bibfnamefont {D.~L.}\ \bibnamefont
  {Mills}},\ }\href@noop {} {\bibfield  {journal} {\bibinfo  {journal} {Phys.
  Rev. B},\ }\textbf {\bibinfo {volume} {68}},\ \bibinfo {pages} {224414}
  (\bibinfo {year} {2003})}\BibitemShut {NoStop}%
\bibitem [{\citenamefont {Lounis}\ \emph {et~al.}(2010)\citenamefont {Lounis},
  \citenamefont {Costa}, \citenamefont {Muniz},\ and\ \citenamefont
  {Mills}}]{Lounis+PRL10}%
  \BibitemOpen
  \bibfield  {author} {\bibinfo {author} {\bibfnamefont {S.}~\bibnamefont
  {Lounis}}, \bibinfo {author} {\bibfnamefont {A.~T.}\ \bibnamefont {Costa}},
  \bibinfo {author} {\bibfnamefont {R.~B.}\ \bibnamefont {Muniz}}, \ and\
  \bibinfo {author} {\bibfnamefont {D.~L.}\ \bibnamefont {Mills}},\ }\href@noop
  {} {\bibfield  {journal} {\bibinfo  {journal} {Phys. Rev. Lett.},\ }\textbf
  {\bibinfo {volume} {105}},\ \bibinfo {pages} {187205} (\bibinfo {year}
  {2010})}\BibitemShut {NoStop}%
\bibitem [{\citenamefont {Lounis}\ \emph {et~al.}(2011)\citenamefont {Lounis},
  \citenamefont {Costa}, \citenamefont {Muniz},\ and\ \citenamefont
  {Mills}}]{Lounis+PRB11}%
  \BibitemOpen
  \bibfield  {author} {\bibinfo {author} {\bibfnamefont {S.}~\bibnamefont
  {Lounis}}, \bibinfo {author} {\bibfnamefont {A.~T.}\ \bibnamefont {Costa}},
  \bibinfo {author} {\bibfnamefont {R.~B.}\ \bibnamefont {Muniz}}, \ and\
  \bibinfo {author} {\bibfnamefont {D.~L.}\ \bibnamefont {Mills}},\ }\href@noop
  {} {\bibfield  {journal} {\bibinfo  {journal} {Phys. Rev. B},\ }\textbf
  {\bibinfo {volume} {83}},\ \bibinfo {pages} {035109} (\bibinfo {year}
  {2011})}\BibitemShut {NoStop}%
\bibitem [{\citenamefont {Lorente}\ and\ \citenamefont
  {Gauyacq}(2009)}]{LorenteGauyacqPRL09}%
  \BibitemOpen
  \bibfield  {author} {\bibinfo {author} {\bibfnamefont {N.}~\bibnamefont
  {Lorente}}\ and\ \bibinfo {author} {\bibfnamefont {J.-P.}\ \bibnamefont
  {Gauyacq}},\ }\href@noop {} {\bibfield  {journal} {\bibinfo  {journal} {Phys.
  Rev. Lett.},\ }\textbf {\bibinfo {volume} {103}},\ \bibinfo {pages} {176601}
  (\bibinfo {year} {2009})}\BibitemShut {NoStop}%
\bibitem [{\citenamefont {Persson}(2009)}]{PerssonPRL09}%
  \BibitemOpen
  \bibfield  {author} {\bibinfo {author} {\bibfnamefont {M.}~\bibnamefont
  {Persson}},\ }\href@noop {} {\bibfield  {journal} {\bibinfo  {journal} {Phys.
  Rev. Lett.},\ }\textbf {\bibinfo {volume} {103}},\ \bibinfo {pages} {050801}
  (\bibinfo {year} {2009})}\BibitemShut {NoStop}%
\bibitem [{\citenamefont {Fern\'andez-Rossier}(2009)}]{Fernandez-RossierPRL09}%
  \BibitemOpen
  \bibfield  {author} {\bibinfo {author} {\bibfnamefont {J.}~\bibnamefont
  {Fern\'andez-Rossier}},\ }\href@noop {} {\bibfield  {journal} {\bibinfo
  {journal} {Phys. Rev. Lett.},\ }\textbf {\bibinfo {volume} {102}},\ \bibinfo
  {pages} {256802} (\bibinfo {year} {2009})}\BibitemShut {NoStop}%
\bibitem [{\citenamefont {Fransson}(2009)}]{FranssonNL09}%
  \BibitemOpen
  \bibfield  {author} {\bibinfo {author} {\bibfnamefont {J.}~\bibnamefont
  {Fransson}},\ }\href@noop {} {\bibfield  {journal} {\bibinfo  {journal} {Nano
  Lett.},\ }\textbf {\bibinfo {volume} {9}},\ \bibinfo {pages} {2414} (\bibinfo
  {year} {2009})}\BibitemShut {NoStop}%
\bibitem [{\citenamefont {Hurley}\ \emph {et~al.}(2011)\citenamefont {Hurley},
  \citenamefont {Baadji},\ and\ \citenamefont {Sanvito}}]{Hurley+PRB11}%
  \BibitemOpen
  \bibfield  {author} {\bibinfo {author} {\bibfnamefont {A.}~\bibnamefont
  {Hurley}}, \bibinfo {author} {\bibfnamefont {N.}~\bibnamefont {Baadji}}, \
  and\ \bibinfo {author} {\bibfnamefont {S.}~\bibnamefont {Sanvito}},\
  }\href@noop {} {\bibfield  {journal} {\bibinfo  {journal} {Phys. Rev. B},\
  }\textbf {\bibinfo {volume} {84}},\ \bibinfo {pages} {035427} (\bibinfo
  {year} {2011})}\BibitemShut {NoStop}%
\bibitem [{\citenamefont {Hurley}\ \emph {et~al.}(2012)\citenamefont {Hurley},
  \citenamefont {Baadji},\ and\ \citenamefont {Sanvito}}]{Hurley+PRB12}%
  \BibitemOpen
  \bibfield  {author} {\bibinfo {author} {\bibfnamefont {A.}~\bibnamefont
  {Hurley}}, \bibinfo {author} {\bibfnamefont {N.}~\bibnamefont {Baadji}}, \
  and\ \bibinfo {author} {\bibfnamefont {S.}~\bibnamefont {Sanvito}},\
  }\href@noop {} {\bibfield  {journal} {\bibinfo  {journal} {Phys. Rev. B},\
  }\textbf {\bibinfo {volume} {86}},\ \bibinfo {pages} {125411} (\bibinfo
  {year} {2012})}\BibitemShut {NoStop}%
\bibitem [{\citenamefont {Tersoff}\ and\ \citenamefont
  {Hamann}(1983)}]{TersoffHamannPRL83}%
  \BibitemOpen
  \bibfield  {author} {\bibinfo {author} {\bibfnamefont {J.}~\bibnamefont
  {Tersoff}}\ and\ \bibinfo {author} {\bibfnamefont {D.~R.}\ \bibnamefont
  {Hamann}},\ }\href@noop {} {\bibfield  {journal} {\bibinfo  {journal} {Phys.
  Rev. Lett.},\ }\textbf {\bibinfo {volume} {50}},\ \bibinfo {pages} {1998}
  (\bibinfo {year} {1983})}\BibitemShut {NoStop}%
\bibitem [{\citenamefont {Balashov}\ \emph {et~al.}(2008)\citenamefont
  {Balashov}, \citenamefont {Tak\'acs}, \citenamefont {D\"ane}, \citenamefont
  {Ernst}, \citenamefont {Bruno},\ and\ \citenamefont
  {Wulfhekel}}]{Balashov+PRB08}%
  \BibitemOpen
  \bibfield  {author} {\bibinfo {author} {\bibfnamefont {T.}~\bibnamefont
  {Balashov}}, \bibinfo {author} {\bibfnamefont {A.~F.}\ \bibnamefont
  {Tak\'acs}}, \bibinfo {author} {\bibfnamefont {M.}~\bibnamefont {D\"ane}},
  \bibinfo {author} {\bibfnamefont {A.}~\bibnamefont {Ernst}}, \bibinfo
  {author} {\bibfnamefont {P.}~\bibnamefont {Bruno}}, \ and\ \bibinfo {author}
  {\bibfnamefont {W.}~\bibnamefont {Wulfhekel}},\ }\href@noop {} {\bibfield
  {journal} {\bibinfo  {journal} {Phys. Rev. B},\ }\textbf {\bibinfo {volume}
  {78}},\ \bibinfo {pages} {174404} (\bibinfo {year} {2008})}\BibitemShut
  {NoStop}%
\bibitem [{\citenamefont {Baym}\ and\ \citenamefont
  {Kadanoff}(1961)}]{BaymKadanoffPR61}%
  \BibitemOpen
  \bibfield  {author} {\bibinfo {author} {\bibfnamefont {G.}~\bibnamefont
  {Baym}}\ and\ \bibinfo {author} {\bibfnamefont {L.~P.}\ \bibnamefont
  {Kadanoff}},\ }\href@noop {} {\bibfield  {journal} {\bibinfo  {journal}
  {Phys. Rev.},\ }\textbf {\bibinfo {volume} {124}},\ \bibinfo {pages} {287}
  (\bibinfo {year} {1961})}\BibitemShut {NoStop}%
\bibitem [{\citenamefont {Kanamori}(1963)}]{KanamoriPTP63}%
  \BibitemOpen
  \bibfield  {author} {\bibinfo {author} {\bibnamefont {Kanamori}},\
  }\href@noop {} {\bibfield  {journal} {\bibinfo  {journal} {Prog. Theo.
  Phys.},\ }\textbf {\bibinfo {volume} {30}},\ \bibinfo {pages} {275} (\bibinfo
  {year} {1963})}\BibitemShut {NoStop}%
\bibitem [{Note1()}]{Note1}%
  \BibitemOpen
  \bibinfo {note} {$\Sigma $ is given as a convolution $G(T-U)$. Since $T = U +
  U\chi U$, $\Sigma $ simplifies to $GU\chi U$ where $\chi $ is the transverse
  magnetic response function.}\BibitemShut {Stop}%
\bibitem [{\citenamefont {Wang}\ and\ \citenamefont
  {Scalapino}(1968)}]{WangScalapinoPR68}%
  \BibitemOpen
  \bibfield  {author} {\bibinfo {author} {\bibfnamefont {Y.-L.}\ \bibnamefont
  {Wang}}\ and\ \bibinfo {author} {\bibfnamefont {D.~J.}\ \bibnamefont
  {Scalapino}},\ }\href@noop {} {\bibfield  {journal} {\bibinfo  {journal}
  {Phys. Rev.},\ }\textbf {\bibinfo {volume} {175}},\ \bibinfo {pages} {734}
  (\bibinfo {year} {1968})}\BibitemShut {NoStop}%
\bibitem [{\citenamefont {Appelbaum}\ and\ \citenamefont
  {Brinkman}(1969)}]{AppelbaumBrinkmanPR69}%
  \BibitemOpen
  \bibfield  {author} {\bibinfo {author} {\bibfnamefont {J.~A.}\ \bibnamefont
  {Appelbaum}}\ and\ \bibinfo {author} {\bibfnamefont {W.~F.}\ \bibnamefont
  {Brinkman}},\ }\href@noop {} {\bibfield  {journal} {\bibinfo  {journal}
  {Phys. Rev.},\ }\textbf {\bibinfo {volume} {183}},\ \bibinfo {pages} {553}
  (\bibinfo {year} {1969})}\BibitemShut {NoStop}%
\bibitem [{\citenamefont {Edwards}\ and\ \citenamefont
  {Hertz}(1973){\natexlab{a}}}]{EdwardsHertzJPF73-I}%
  \BibitemOpen
  \bibfield  {author} {\bibinfo {author} {\bibfnamefont {D.~M.}\ \bibnamefont
  {Edwards}}\ and\ \bibinfo {author} {\bibfnamefont {J.~A.}\ \bibnamefont
  {Hertz}},\ }\href@noop {} {\bibfield  {journal} {\bibinfo  {journal} {Journal
  of Physics F: Metal Physics},\ }\textbf {\bibinfo {volume} {3}},\ \bibinfo
  {pages} {2174} (\bibinfo {year} {1973}{\natexlab{a}})}\BibitemShut {NoStop}%
\bibitem [{\citenamefont {Edwards}\ and\ \citenamefont
  {Hertz}(1973){\natexlab{b}}}]{EdwardsHertzJPF73-II}%
  \BibitemOpen
  \bibfield  {author} {\bibinfo {author} {\bibfnamefont {D.~M.}\ \bibnamefont
  {Edwards}}\ and\ \bibinfo {author} {\bibfnamefont {J.~A.}\ \bibnamefont
  {Hertz}},\ }\href@noop {} {\bibfield  {journal} {\bibinfo  {journal} {Journal
  of Physics F: Metal Physics},\ }\textbf {\bibinfo {volume} {3}},\ \bibinfo
  {pages} {2191} (\bibinfo {year} {1973}{\natexlab{b}})}\BibitemShut {NoStop}%
\bibitem [{\citenamefont {Celasco}\ and\ \citenamefont
  {Corrias}(1976)}]{CelascoCorriasNC76}%
  \BibitemOpen
  \bibfield  {author} {\bibinfo {author} {\bibfnamefont {M.}~\bibnamefont
  {Celasco}}\ and\ \bibinfo {author} {\bibfnamefont {M.}~\bibnamefont
  {Corrias}},\ }\href@noop {} {\bibfield  {journal} {\bibinfo  {journal} {Nuovo
  Cimento},\ }\textbf {\bibinfo {volume} {33}},\ \bibinfo {pages} {807}
  (\bibinfo {year} {1976})}\BibitemShut {NoStop}%
\bibitem [{\citenamefont {Hong}\ and\ \citenamefont
  {Mills}(1999)}]{HongMillsPRB99}%
  \BibitemOpen
  \bibfield  {author} {\bibinfo {author} {\bibfnamefont {J.}~\bibnamefont
  {Hong}}\ and\ \bibinfo {author} {\bibfnamefont {D.~L.}\ \bibnamefont
  {Mills}},\ }\href@noop {} {\bibfield  {journal} {\bibinfo  {journal} {Phys.
  Rev. B},\ }\textbf {\bibinfo {volume} {59}},\ \bibinfo {pages} {13840}
  (\bibinfo {year} {1999})}\BibitemShut {NoStop}%
\bibitem [{\citenamefont {Zhukov}\ \emph {et~al.}(2004)\citenamefont {Zhukov},
  \citenamefont {Chulkov},\ and\ \citenamefont {Echenique}}]{Zhukov+PRL04}%
  \BibitemOpen
  \bibfield  {author} {\bibinfo {author} {\bibfnamefont {V.~P.}\ \bibnamefont
  {Zhukov}}, \bibinfo {author} {\bibfnamefont {E.~V.}\ \bibnamefont {Chulkov}},
  \ and\ \bibinfo {author} {\bibfnamefont {P.~M.}\ \bibnamefont {Echenique}},\
  }\href@noop {} {\bibfield  {journal} {\bibinfo  {journal} {Phys. Rev.
  Lett.},\ }\textbf {\bibinfo {volume} {93}},\ \bibinfo {pages} {096401}
  (\bibinfo {year} {2004})}\BibitemShut {NoStop}%
\bibitem [{\citenamefont {Zhukov}\ \emph {et~al.}(2006)\citenamefont {Zhukov},
  \citenamefont {Chulkov},\ and\ \citenamefont {Echenique}}]{Zhukov+PRB06}%
  \BibitemOpen
  \bibfield  {author} {\bibinfo {author} {\bibfnamefont {V.~P.}\ \bibnamefont
  {Zhukov}}, \bibinfo {author} {\bibfnamefont {E.~V.}\ \bibnamefont {Chulkov}},
  \ and\ \bibinfo {author} {\bibfnamefont {P.~M.}\ \bibnamefont {Echenique}},\
  }\href@noop {} {\bibfield  {journal} {\bibinfo  {journal} {Phys. Rev. B},\
  }\textbf {\bibinfo {volume} {73}},\ \bibinfo {pages} {125105} (\bibinfo
  {year} {2006})}\BibitemShut {NoStop}%
\bibitem [{\citenamefont {M\"uller}(2011)}]{MuellerMasterThesis11}%
  \BibitemOpen
  \bibfield  {author} {\bibinfo {author} {\bibfnamefont {M.~C. T.~D.}\
  \bibnamefont {M\"uller}},\ }\emph {\bibinfo {title} {Electron-Magnon
  Interaction in $GT$ Approximation}},\ \href@noop {} {Master's thesis},\
  \bibinfo  {school} {RWTH Aachen University} (\bibinfo {year}
  {2011})\BibitemShut {NoStop}%
\bibitem [{Note2()}]{Note2}%
  \BibitemOpen
  \bibinfo {note} {In Refs.~[\protect \rev@citealp
  {Zhukov+PRL04,Zhukov+PRB06,MuellerMasterThesis11}], the screened interaction,
  $W$, as calculated in $GW$ was used in evaluating $T$ instead of $U$.
  Romaniello \protect \textit {et al}.~\cite {Romaniello+PRB12} discusses the
  different forms of $T$ depending on the strength of screening.}\BibitemShut
  {Stop}%
\bibitem [{\citenamefont {\ifmmode \mbox{\c{S}}\else \c{S}\fi{}a\ifmmode
  \mbox{\c{s}}\else \c{s}\fi{}\ifmmode \imath \else \i
  \fi{}o\ifmmode~\breve{g}\else \u{g}\fi{}lu}\ \emph
  {et~al.}(2010)\citenamefont {\ifmmode \mbox{\c{S}}\else \c{S}\fi{}a\ifmmode
  \mbox{\c{s}}\else \c{s}\fi{}\ifmmode \imath \else \i
  \fi{}o\ifmmode~\breve{g}\else \u{g}\fi{}lu}, \citenamefont {Schindlmayr},
  \citenamefont {Friedrich}, \citenamefont {Freimuth},\ and\ \citenamefont
  {Bl\"ugel}}]{Sasioglu+PRB10}%
  \BibitemOpen
  \bibfield  {author} {\bibinfo {author} {\bibfnamefont {E.}~\bibnamefont
  {\ifmmode \mbox{\c{S}}\else \c{S}\fi{}a\ifmmode \mbox{\c{s}}\else
  \c{s}\fi{}\ifmmode \imath \else \i \fi{}o\ifmmode~\breve{g}\else
  \u{g}\fi{}lu}}, \bibinfo {author} {\bibfnamefont {A.}~\bibnamefont
  {Schindlmayr}}, \bibinfo {author} {\bibfnamefont {C.}~\bibnamefont
  {Friedrich}}, \bibinfo {author} {\bibfnamefont {F.}~\bibnamefont {Freimuth}},
  \ and\ \bibinfo {author} {\bibfnamefont {S.}~\bibnamefont {Bl\"ugel}},\
  }\href@noop {} {\bibfield  {journal} {\bibinfo  {journal} {Phys. Rev. B},\
  }\textbf {\bibinfo {volume} {81}},\ \bibinfo {pages} {054434} (\bibinfo
  {year} {2010})}\BibitemShut {NoStop}%
\bibitem [{\citenamefont {Karlsson}\ and\ \citenamefont
  {Aryasetiawan}(2004)}]{KarlssonAryasetiawanIJMPB04}%
  \BibitemOpen
  \bibfield  {author} {\bibinfo {author} {\bibfnamefont {K.}~\bibnamefont
  {Karlsson}}\ and\ \bibinfo {author} {\bibfnamefont {F.}~\bibnamefont
  {Aryasetiawan}},\ }\href@noop {} {\bibfield  {journal} {\bibinfo  {journal}
  {International J. of Mod. Phys. B},\ }\textbf {\bibinfo {volume} {18}},\
  \bibinfo {pages} {1055} (\bibinfo {year} {2004})}\BibitemShut {NoStop}%
\bibitem [{\citenamefont {Brandt}(1971)}]{Brandt71}%
  \BibitemOpen
  \bibfield  {author} {\bibinfo {author} {\bibfnamefont {U.}~\bibnamefont
  {Brandt}},\ }\href@noop {} {\bibfield  {journal} {\bibinfo  {journal} {Z.
  Phys.},\ }\textbf {\bibinfo {volume} {244}},\ \bibinfo {pages} {217}
  (\bibinfo {year} {1971})}\BibitemShut {NoStop}%
\bibitem [{\citenamefont {Papanikolaou}\ \emph {et~al.}(2002)\citenamefont
  {Papanikolaou}, \citenamefont {Zeller},\ and\ \citenamefont
  {Dederichs}}]{KKR}%
  \BibitemOpen
  \bibfield  {author} {\bibinfo {author} {\bibfnamefont {N.}~\bibnamefont
  {Papanikolaou}}, \bibinfo {author} {\bibfnamefont {R.}~\bibnamefont
  {Zeller}}, \ and\ \bibinfo {author} {\bibfnamefont {P.~H.}\ \bibnamefont
  {Dederichs}},\ }\href@noop {} {\bibfield  {journal} {\bibinfo  {journal}
  {Journal of Physics: Condensed Matter},\ }\textbf {\bibinfo {volume} {14}},\
  \bibinfo {pages} {2799} (\bibinfo {year} {2002})}\BibitemShut {NoStop}%
\bibitem [{\citenamefont {Vosko}\ \emph {et~al.}(1980)\citenamefont {Vosko},
  \citenamefont {Wilk},\ and\ \citenamefont {Nusair}}]{VoskoWilkNusairCJP80}%
  \BibitemOpen
  \bibfield  {author} {\bibinfo {author} {\bibfnamefont {S.~H.}\ \bibnamefont
  {Vosko}}, \bibinfo {author} {\bibfnamefont {L.}~\bibnamefont {Wilk}}, \ and\
  \bibinfo {author} {\bibfnamefont {M.}~\bibnamefont {Nusair}},\ }\href@noop {}
  {\bibfield  {journal} {\bibinfo  {journal} {Canadian Journal of Physics},\
  }\textbf {\bibinfo {volume} {58}},\ \bibinfo {pages} {1200} (\bibinfo {year}
  {1980})}\BibitemShut {NoStop}%
\bibitem [{\citenamefont {Lounis}\ \emph {et~al.}(2006)\citenamefont {Lounis},
  \citenamefont {Mavropoulos}, \citenamefont {Dederichs},\ and\ \citenamefont
  {Bl\"ugel}}]{Lounis+PRB06}%
  \BibitemOpen
  \bibfield  {author} {\bibinfo {author} {\bibfnamefont {S.}~\bibnamefont
  {Lounis}}, \bibinfo {author} {\bibfnamefont {P.}~\bibnamefont {Mavropoulos}},
  \bibinfo {author} {\bibfnamefont {P.~H.}\ \bibnamefont {Dederichs}}, \ and\
  \bibinfo {author} {\bibfnamefont {S.}~\bibnamefont {Bl\"ugel}},\ }\href@noop
  {} {\bibfield  {journal} {\bibinfo  {journal} {Phys. Rev. B},\ }\textbf
  {\bibinfo {volume} {73}},\ \bibinfo {pages} {195421} (\bibinfo {year}
  {2006})}\BibitemShut {NoStop}%
\bibitem [{\citenamefont {Limot}\ \emph {et~al.}(2005)\citenamefont {Limot},
  \citenamefont {Pehlke}, \citenamefont {Kr\"oger},\ and\ \citenamefont
  {Berndt}}]{Limot+PRL05}%
  \BibitemOpen
  \bibfield  {author} {\bibinfo {author} {\bibfnamefont {L.}~\bibnamefont
  {Limot}}, \bibinfo {author} {\bibfnamefont {E.}~\bibnamefont {Pehlke}},
  \bibinfo {author} {\bibfnamefont {J.}~\bibnamefont {Kr\"oger}}, \ and\
  \bibinfo {author} {\bibfnamefont {R.}~\bibnamefont {Berndt}},\ }\href@noop {}
  {\bibfield  {journal} {\bibinfo  {journal} {Phys. Rev. Lett.},\ }\textbf
  {\bibinfo {volume} {94}},\ \bibinfo {pages} {036805} (\bibinfo {year}
  {2005})}\BibitemShut {NoStop}%
\bibitem [{Note3()}]{Note3}%
  \BibitemOpen
  \bibinfo {note} {But spin excitations and Kondo can coexist~\cite
  {Otte+NP08}.}\BibitemShut {Stop}%
\bibitem [{\citenamefont {Khajetoorians}\ \emph {et~al.}(2012)\citenamefont
  {Khajetoorians}, \citenamefont {Wiebe}, \citenamefont {Chilian},
  \citenamefont {Lounis}, \citenamefont {Bl\"ugel},\ and\ \citenamefont
  {Wiesendanger}}]{Khajetoorians+NP12}%
  \BibitemOpen
  \bibfield  {author} {\bibinfo {author} {\bibfnamefont {A.~A.}\ \bibnamefont
  {Khajetoorians}}, \bibinfo {author} {\bibfnamefont {J.}~\bibnamefont
  {Wiebe}}, \bibinfo {author} {\bibfnamefont {B.}~\bibnamefont {Chilian}},
  \bibinfo {author} {\bibfnamefont {S.}~\bibnamefont {Lounis}}, \bibinfo
  {author} {\bibfnamefont {S.}~\bibnamefont {Bl\"ugel}}, \ and\ \bibinfo
  {author} {\bibfnamefont {R.}~\bibnamefont {Wiesendanger}},\ }\href@noop {}
  {\bibfield  {journal} {\bibinfo  {journal} {Nature Phys.},\ }\textbf
  {\bibinfo {volume} {8}},\ \bibinfo {pages} {497} (\bibinfo {year}
  {2012})}\BibitemShut {NoStop}%
\bibitem [{\citenamefont {Jamneala}\ \emph {et~al.}(2000)\citenamefont
  {Jamneala}, \citenamefont {Madhavan}, \citenamefont {Chen},\ and\
  \citenamefont {Crommie}}]{Jamneala+PRB00}%
  \BibitemOpen
  \bibfield  {author} {\bibinfo {author} {\bibfnamefont {T.}~\bibnamefont
  {Jamneala}}, \bibinfo {author} {\bibfnamefont {V.}~\bibnamefont {Madhavan}},
  \bibinfo {author} {\bibfnamefont {W.}~\bibnamefont {Chen}}, \ and\ \bibinfo
  {author} {\bibfnamefont {M.~F.}\ \bibnamefont {Crommie}},\ }\href@noop {}
  {\bibfield  {journal} {\bibinfo  {journal} {Phys. Rev. B},\ }\textbf
  {\bibinfo {volume} {61}},\ \bibinfo {pages} {9990} (\bibinfo {year}
  {2000})}\BibitemShut {NoStop}%
\bibitem [{\citenamefont {Romaniello}\ \emph {et~al.}(2012)\citenamefont
  {Romaniello}, \citenamefont {Bechstedt},\ and\ \citenamefont
  {Reining}}]{Romaniello+PRB12}%
  \BibitemOpen
  \bibfield  {author} {\bibinfo {author} {\bibfnamefont {P.}~\bibnamefont
  {Romaniello}}, \bibinfo {author} {\bibfnamefont {F.}~\bibnamefont
  {Bechstedt}}, \ and\ \bibinfo {author} {\bibfnamefont {L.}~\bibnamefont
  {Reining}},\ }\href@noop {} {\bibfield  {journal} {\bibinfo  {journal} {Phys.
  Rev. B},\ }\textbf {\bibinfo {volume} {85}},\ \bibinfo {pages} {155131}
  (\bibinfo {year} {2012})}\BibitemShut {NoStop}%
\bibitem [{\citenamefont {Otte}\ \emph {et~al.}(2008)\citenamefont {Otte},
  \citenamefont {Ternes}, \citenamefont {von Bergmann}, \citenamefont {Loth},
  \citenamefont {Brune}, \citenamefont {Lutz}, \citenamefont {Hirjibehedin},\
  and\ \citenamefont {Heinrich}}]{Otte+NP08}%
  \BibitemOpen
  \bibfield  {author} {\bibinfo {author} {\bibfnamefont {A.~F.}\ \bibnamefont
  {Otte}}, \bibinfo {author} {\bibfnamefont {M.}~\bibnamefont {Ternes}},
  \bibinfo {author} {\bibfnamefont {K.}~\bibnamefont {von Bergmann}}, \bibinfo
  {author} {\bibfnamefont {S.}~\bibnamefont {Loth}}, \bibinfo {author}
  {\bibfnamefont {H.}~\bibnamefont {Brune}}, \bibinfo {author} {\bibfnamefont
  {C.~P.}\ \bibnamefont {Lutz}}, \bibinfo {author} {\bibfnamefont {C.~F.}\
  \bibnamefont {Hirjibehedin}}, \ and\ \bibinfo {author} {\bibfnamefont
  {A.~J.}\ \bibnamefont {Heinrich}},\ }\href@noop {} {\bibfield  {journal}
  {\bibinfo  {journal} {Nature Phys.},\ }\textbf {\bibinfo {volume} {4}},\
  \bibinfo {pages} {847} (\bibinfo {year} {2008})}\BibitemShut {NoStop}%
\end{thebibliography}%

\end{document}